\documentclass[aps,prb,twocolumn,superscriptaddress,groupedaddress]{revtex4}
\usepackage[english]{babel}
\usepackage{epstopdf}
\usepackage{pdfpages}
\usepackage{float}
\usepackage{enumerate}

\usepackage{amsmath} % AMS Math Package
\usepackage{dsfont}
\usepackage{cancel}
\usepackage{amsthm} % Theorem Formatting
\usepackage{amssymb}	% Math symbols such as \mathbb
\usepackage{empheq}
\usepackage[absolute,overlay]{textpos}
\usepackage{xcolor}

\allowdisplaybreaks

\let\baraccent=\= % rename builtin command \= to \baraccent
\renewcommand{\=}[1]{\stackrel{#1}{=}} % for putting numbers above =

\begin{document}

\title{ Universal quantum criticality in static and Floquet-Majorana chains}

\date{\today}

\author{Paolo Molignini}
%\email{molignini@itp.phys.ethz.ch}
\affiliation{Institute for Theoretical Physics, ETH Z\"{u}rich, 8093 Zurich, Switzerland}
\author{Wei Chen}
\affiliation{Institute for Theoretical Physics, ETH Z\"{u}rich, 8093 Zurich, Switzerland}
\affiliation{Department of Physics, PUC-Rio, 22451-900 Rio de Janeiro, Brazil}
\author{R. Chitra}
\affiliation{Institute for Theoretical Physics, ETH Z\"{u}rich, 8093 Zurich, Switzerland}

\begin{abstract}

 The topological phase transitions in static and periodically driven Kitaev chains are investigated by means of a renormalization group (RG) approach.  These transitions, across which  the numbers of static or Floquet Majorana edge modes  change,  are accompanied by divergences of the  relevant Berry connections. These divergences at certain high symmetry points in momentum space form
 the basis of the RG approach, through which  topological phase boundaries  are identified  as a function of  system parameters.  We also introduce several aspects to characterize the quantum criticality of the topological phase transitions in both static and Floquet systems: a correlation function that measures the overlap of Majorana-Wannier functions,  the decay length of the Majorana edge mode and a scaling law relating the critical exponents.  These indicate a common universal critical behavior for topological phase transitions, in both static and periodically driven chains.
For the latter, the RG flows  additionally display intriguing  features  related to gap closures  at non-high symmetry points due to momentarily frozen dynamics.
\end{abstract}

\maketitle

\section{Introduction}

%General topology intro
The discovery of topological order has enriched the theory of phase transitions with a new fundamental paradigm~\citep{Wen:1990}. 
Contrary to the traditional Landau formalism based on spontaneous symmetry breaking~\citep{Landau,Miransky-book}, topologically ordered systems are not described by local order parameters but by robust ground state degeneracy, quantized geometric phases~\citep{Thouless:1982, Wen:1989, Qi:2008} and often long-range quantum entanglement~\citep{Wen:1995, XieChen:2010,Fidkowski:2010}. 
Moreover, bulk-edge correspondences lead to a wide variety of edge states in topological systems~\citep{Wen:1991, Kane-Mele2005}.   
Topological systems  can  exhibit charge fractionalization,  as well as excitations  with exotic abelian and  non-abelian  statistics~\citep{Goldman:1995, dePicciotto:1997, Martin:2004, Moore, Kitaev2001, Mong:2014}.  These can be harnessed for revolutionary applications, such as spintronics with edge currents~\citep{Sato-Spintronics-book}, topological quantum memory devices with highly-entangled matter~\citep{Dennis:2002}, and, most notably, fault-tolerant topological quantum computation~\citep{Kitaev:2003, Nayak:2008, Mong:2014}.

Recently, the exploration of topological order has been extended to Floquet systems, where it can be generated through driving in otherwise topologically trivial systems.
Prominent examples are Floquet topological insulators, and Floquet topological superconductors that host
Floquet-Majorana  modes ~\citep{Lindner, Kitagawa, Liu, Cayssol:2013, Harper:2017, Roy:2017, Yao:2017}.
Floquet systems in one dimension  often exhibit a tunable topology wherein, the number of  edge modes can be systematically  increased  by manipulating the intensity or the frequency of the  drive~\citep{Liu, Thakurathi, Molignini:2017}.   
This results in a series of gap closures in the quasienergy spectrum which signal out-of-equilibrium Floquet topological phase transitions (TPT) between topologically inequivalent phases~\citep{Graf:2013,Molignini:2017}.

%top RG intro
The primary focus of literature has been on topological classifications of these phases based on symmetries rather than the nature of the transitions themselves~\citep{ChiuReview:2016}.
Recently, a renormalization group approach was proposed to study the nature of TPTs  in static  systems~\citep{Chen:2016,Chen-Sigrist:2016}.
It essentially exploits the idea that, since the topological invariant generally takes the form of an integration over a certain curvature function, TPTs can be identified through appropriate deformations of the curvature function, analogous to stretching a messy string to reveal the number of knots it contains.
This method, termed the curvature renormalization group (CRG) approach, has been successful in describing TPTs in a variety of interacting and noninteracting static models~\citep{Chen:2016,Chen-Sigrist:2016,Kourtis:2017,Chen:2018}. 

%In this article
In this article, we extend the CRG scheme to Floquet systems.  To benchmark our method, we   study TPTs in  both the static and  periodically driven  Kitaev chains.
The scheme is based on deformations of the Berry connection of the appropriate Bloch or Floquet-Bloch eigenstate of the static or effective Floquet hamiltonian, which plays the role of the curvature function in this problem.
Though winding numbers based on this curvature function incompletely reproduce the TPTs in Floquet systems~\citep{Thakurathi, Rudner:2013},  the curvature function always diverges at certain high symmetry points (HSPs) in momentum space as one approaches the TPTs.
We will show that the latter  feature  determines the critical points of our CRG  and suffices to obtain the full topological phase diagram in an extremely simplified manner.  
We find that TPTs  across which the number of edge Majorana modes change by one are characterized by certain universal features:  divergence of the Majorana-Wannier state correlation length and a scaling law that constrains the critical exponents.
Additionally, the fixed points of the CRG flow reveal another, more subtle type of instability, where the driving-induced dynamics is frozen revealing
 minimal correlations.
Intriguingly, some of these fixed lines are  also associated with gap closings in the quasi-energy spectrum  opening up the potential for new kinds of TPTs in driven systems.

%Structure of the article
The article is structured  as follows.
In Sec.~\ref{sec:generic_critical_behavior}, we present an overview of  the CRG method  based on the Berry connection function. 
In Sec.~\ref{sec:static_kitaev_chain}, we present an illustration of the method by applying it to the static Kitaev chain for fermions or, equivalently, the XY spin-$\frac12$ chain in a transverse magnetic field. 
In Sec.~\ref{sec:periodically_driven_Kitaev}, we apply the methodology to the periodically driven Kitaev chain and present a general analysis of the Floquet TPTs which exist in these systems.
Finally, Sec.~\ref{sec:conclusions} summarizes the main results of the article and offers a glimpse in open questions and possible future directions.

%%%%%%%%%%%%%%%%%%%%%%%%
% SECTION: CRG APPROACH
%%%%%%%%%%%%%%%%%%%%%%%%
\section{Topological phase transitions and a renormalization group approach\label{sec:generic_critical_behavior}}
Here, we briefly review the CRG approach expostulated in \citep{Chen:2016, Chen-Sigrist:2016, Chen:2017}, which is designed to capture the critical behaviour of a  static system close to a TPT.
We consider a topological system, whose critical behavior at the TPTs is  driven by a set of tuning parameters ${\bf M}=\left(M_{1},M_{2},...\right)$  in the Hamiltonian.
For example, if we consider the XY spin chain in a transverse field, the tuning parameters are  the magnetic field and the anisotropy, or in the equivalent Kitaev chain model, the parameters are the chemical potential and the pairing gap.

We denote by $F(k, {\bf M})$,  the generic  curvature function in one dimension that is  synonymous  with  the notion of  the curvature of a closed string  whose integral counts the number of knots it contains. We will elaborate on the relation between $F$ and  the Berry connection in $k$-space in Sec.~\ref{sec:static_Kitaev_chain}.  
For static systems, this curvature function determines the  topological properties of the system  via the winding number defined by 
\begin{equation}
W = \int_{-\pi}^{\pi} \frac{\mathrm{d}k}{2\pi}F(k, {\bf M}).
\label{winding_number_definition}
\end{equation}
Phases with different $W$ are separated by TPTs.  
As discussed extensively in Refs.~\citep{Chen:2016,Chen-Sigrist:2016,Chen:2017}, near HSPs $k_0$ of the underlying lattice, $F(k, {\bf M})$ typically displays the Ornstein-Zernike form
\begin{eqnarray}
F(k_{0}+\delta k,{\bf M})=\frac{F(k_{0},{\bf M})}{1+\xi_{k_0}^{2}\delta k^{2}}\;.
\label{Ornstein_Zernike_fit}
\end{eqnarray}
Let ${\bf M}_c$ denote the critical point where the system undergoes a TPT associated with a gap closure at a certain $k_0$.

When ${\bf M}\rightarrow{\bf M}_{c}$, the length scale $\xi_{k_0} \to \infty$  resulting in  a narrowing of   the Lorentzian  of Eq.~(\ref{Ornstein_Zernike_fit})  and  a divergence of the curvature function as ${\bf M}_c$ is approached from below or above:
\begin{eqnarray}
&&\lim_{{\bf M}\rightarrow{\bf M}_{c}^{+}}F(k_{0},{\bf M}) = -\lim_{{\bf M}\rightarrow{\bf M}_{c}^{-}}F(k_{0},{\bf M}) = \pm\infty \, .
\label{Fk0_xi_divergence}
\end{eqnarray}

Close to the TPT,  we expect the following divergent behavior:
\begin{eqnarray}
F(k_{0},{\bf M})\propto |{\bf M}-{\bf M}_{c}|^{-\gamma}\;,\;\;\;\xi_{k_0}\propto |{\bf M}-{\bf M}_{c}|^{-\nu}\;.
\label{Fk0_xi_critical_exponent}
\end{eqnarray}
 with exponents $\gamma$ and $\nu$ characterizing the underlying TPT. The conservation of the topological invariant as ${\bf M}\rightarrow{\bf M}_{c}$, however, imposes  a scaling law $\gamma=\nu$~\citep{Chen:2017}.
 
The exponents $\nu$  and $\gamma$ are synonymous to those assigned to correlation length and susceptibility exponents within the Landau paradigm.    
To see this, we consider the Fourier transform of the curvature function:
\begin{eqnarray}
\lambda_{R}=\int_{-\pi}^{\pi}\frac{\mathrm{d}k}{2\pi}e^{ikR}F(k,{\bf M})\, .
\label{lambdaR_Fourier_trans}
\end{eqnarray}
The quantity $\lambda_R$ yields  a { \it{ Majorana-Wannier state correlation function}} that exemplifies the proximity of the system to a TPT.  Inserting  Eq.~(\ref{Ornstein_Zernike_fit}) into Eq.~({\ref{lambdaR_Fourier_trans}}), we
see that the correlation function decays exponentially $\lambda_R \propto  \exp(-R/\xi_{k_0})$. This justifies the notion of $\xi_{k_0}$ as the correlation length of the TPT with the associated critical exponents $\nu$. 
On the other hand, the curvature function at HSP is the integration of the correlation function $\int \lambda_{R}\mathrm{d}R=F(k_{0}=0,{\bf M})$, which plays the role of the susceptibility in the Landau order parameter paradigm, and hence the exponent $\gamma$ is assigned.

\begin{figure}[ht]
\centering
\includegraphics[width=0.8\columnwidth]{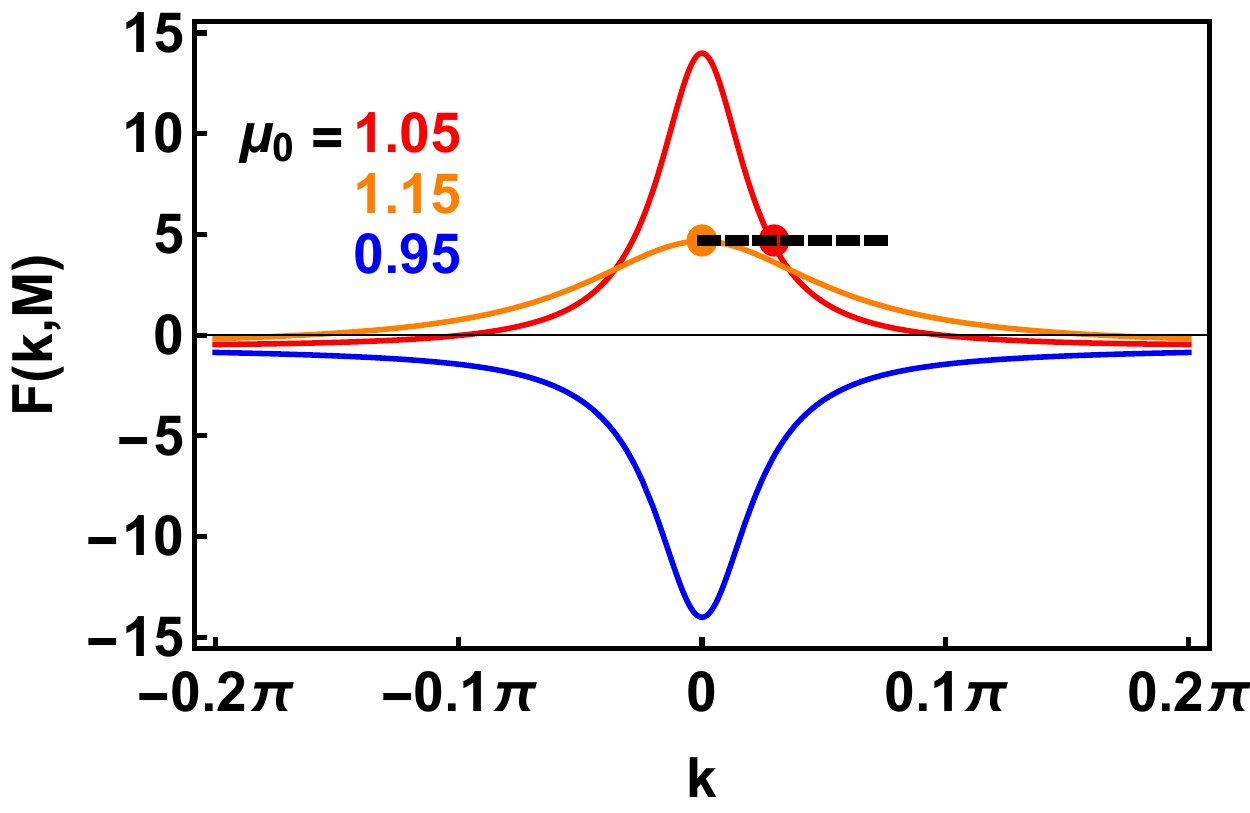}
\caption{The curvature function $F(k,{\bf M})$ near the HSP $k_{0}=0$ for the static Kitaev chain plotted for several values of $\mu_0$ at $\Delta=0.7$. The critical point is located at $\mu_{0}=1$. As approaching the critical point, the curvature function develops a divergence at the HSP (compare orange line and red line), and the divergence flips sign as the system crosses the critical point (compare red line and blue line). The CRG procedure demands the $F(k_{0}+\delta k,{\bf M})$ (red dot) to be equal to $F(k_{0},{\bf M}^{\prime})$ (orange dot), as indicated by the dashed line, through one obtains the CRG flow ${\bf M}\rightarrow{\bf M}^{\prime}$ along which the divergence is reduced. }
\label{plots-Berry-conn-static}
\end{figure}

Based on the divergence described by (\ref{Fk0_xi_divergence}), the CRG scheme has been proposed to identify the TPTs~\citep{Chen:2016, Chen-Sigrist:2016, Chen:2017}.  
The method is based on the iterative search for the trajectory in parameter space (RG flow) along which the divergence of the curvature function is reduced but the topology remains unchanged.
Under this invariant procedure the system will gradually move away from the critical point and the TPTs can be identified.
The RG flow is obtained by demanding that at a given parameter set ${\bf M}$, the next parameter set ${\bf M}^{\prime}$ in the iteration satisfies
\begin{equation}
F(k_0, {\bf M}^{\prime}) = F(k_0 + \delta k, {\bf M}),
\label{cond-RG-eqn}
\end{equation}
where $k_{0}$ is a HSP and $\delta k$ is a small deviation from it.  It can be rigorously shown that $F(k_{0},{\bf M})$  gradually broadens under this procedure~\citep{Chen:2016}, as schematically depicted
 in Fig.~\ref{plots-Berry-conn-static}.

%%%%%
%%%%%

Writing $\mathrm{d}M_{i} = M_{i}^{\prime} - M_{i}$ and $\delta k^2 \equiv \mathrm{d}l$, and expanding Eq.~(\ref{cond-RG-eqn}) to leading order yields the generic RG equation for the parameters ${\bf M}$:
\begin{equation}
\frac{\mathrm{d}M_{i}}{\mathrm{d} l} = \frac{1}{2} \frac{\partial^2_k F(k, {\bf M}) \big|_{k=k_0}}{\partial_{M_{i}} F(k_{0}, {\bf M})},
\label{generic_RG_equation}
\end{equation}
The critical and fixed points of the flow are defined by the following conditions~\citep{Kourtis:2017}
\begin{eqnarray}
{\rm critical\;point}:&&\left|\frac{d{\bf M}}{dl}\right|\rightarrow\infty,\;{\rm flow\;directs\;away},
\nonumber \\
{\rm\;fixed\;point}:&&\left|\frac{d{\bf M}}{dl}\right|\rightarrow 0,\;{\rm flow\;directs\;into}.
\label{identifying_Mc_Mf}
\end{eqnarray}

The TPTs are signalled by the critical points of the flow that form a $(d_{M}-1)$-dimensional surface in the $d_{M}$-dimensional parameter space.
The fixed points of the flow are instead related to regions of low-correlation where $\xi_{k_0}$ vanishes.
In the following sections, the CRG method will be applied to investigate the TPTs in both the static and the periodically driven Kitaev chain.

%%%%%%%%%%%%%%%%%%%%%%%%%%%%%%%%%
%%%%%%%%%%%%%%%%%%%%%%%%%%%%%%%%%
% SECTION: STATIC KITAEV CHAIN
%%%%%%%%%%%%%%%%%%%%%%%%%%%%%%%%%
%%%%%%%%%%%%%%%%%%%%%%%%%%%%%%%%%
\section{Static Kitaev chain}
\label{sec:static_kitaev_chain}

%%%%%%%%%%%%%%%%%%%%%%%%%%%%%%%%%
% SUBSECTION: Majorana edge modes
%%%%%%%%%%%%%%%%%%%%%%%%%%%%%%%%%
\subsection{Majorana edge modes}

We consider the static Kitaev chain described by the following 1D spinless $p$-wave superconducting Hamiltonian,
\begin{align}
\mathcal{H}_{0} &= \sum_{n=1}^{N-1}\left[  t \left( f_n^{\dagger} f_{n+1} + f_{n+1}^{\dagger} f_n \right) + \Delta \left( f_n f_{n+1} + f_{n+1}^{\dagger} f_n^{\dagger} \right)\right] \nonumber \\
& \qquad -  \mu_0  \sum_{n=1}^{N} (2 f_n^{\dagger} f_n - 1),
\label{ham-fermionic-static}
\end{align}
where $f_n, f_{n}^{(\dagger)}$ are spinless fermionic creation and annihilation operators, $t$ is the hopping, $\Delta$ is the $p$-wave pairing between spinless fermions, and $\mu_0$ is the static chemical potential~\citep{Kitaev2001}. 
Note that the fermionic chain can be exactly mapped to a spin-$\frac{1}{2}$ XY chain in a transverse field via a Jordan-Wigner transformation. Throughout this work we will set $t=1$.  

The Kitaev chain undergoes a  TPT to a  topologically nontrivial phase with edge localized Majorana fermions. This  can be seen by splitting each fermion into two real Majorana fermions
\begin{equation}
w_{2n-1} = f_n + f_n^{\dagger}, \qquad w_{2n} = i(f_n - f_n^{\dagger}),
\label{Majorana_real_space}
\end{equation}
that satisfy the anti-commutation relations $\left\{w_n, w_m \right\} = 2 \delta_{nm}$ and $w_n = w_n^{\dagger}$, \textit{i.e.} they are their own antiparticle.
In this representation, the Hamiltonian reads
\begin{align}
	\label{Hamiltonian_Majorana_static}
	\mathcal{H}^{\text{M}}_{0} &= i \sum_{n=1}^{N-1} \bigg[ \frac{t - \Delta}{2} w_{2n} w_{2n+1}  - \frac{t + \Delta}{2}  	w_{2n-1} w_{2n+2} \bigg] \nonumber \\
	& \quad + i \mu_0 \sum_{n=1}^{N} w_{2n-1} w_{2n} = i \sum_{m,n}^{2N} w_m A_{mn}(t) w_n.
\end{align}
The topologically nontrivial phase with zero-energy  edge localized Majoranas appear for $|\frac{\mu_0}{t}| < 1$\citep{Kitaev2001}. 

To investigate the topology, it is useful to rewrite the Hamiltonian in \eqref{ham-fermionic} in Fourier space %
\begin{align}
\mathcal{H}_o &= 2(t - \mu_0) f_0^{\dagger} f_0 + 2(-t - \mu_0) f_{\pi}^{\dagger} f_{\pi} + \nonumber \\ 
& \quad + \sum_{0<k<\pi} \left( f_{k}^{\dagger} f_{-k} \right) h_k \left( \begin{array}{c} f_k \\ f_{-k}^{\dagger} \end{array} \right)
\label{ham-fermionic-k}
\end{align}
where,
$f_{k} = \frac{1}{\sqrt{N}} \sum_{n=1}^N f_n e^{ikn}$ and 
\begin{equation}
h_k = a_{2,k} \tau^y + a_{3,k} \tau^z =  2\Delta \sin k \tau^y + 2(t \cos k - \mu_0) \tau^z 
\label{Dirac_model_static_Kitaev}
\end{equation}
\noindent
and $\tau^a$ are the standard Pauli matrices.
The BdG-Hamiltonian $h_k$ can be interpreted as a vector in this Pauli space~\citep{Thakurathi} which subtends an angle $\phi_k$ in the $yz$-plane.
Integrating this angle variable over the Brillouin zone yields the desired winding number
\begin{equation}
W = \frac{1}{\mathrm{Vol}(BZ)} \int_{BZ} \mathrm{d} \phi_k. 
\label{winding_number}
\end{equation}
By mapping out $W$ across the parameter space spanned by ${\bf M}=(\mu_0,\Delta)$, one obtains a topologically nontrivial phase with $W=1$ at $|\mu_0|<|t|$, and a trivial phase with $W=0$ at $|\mu_0|>|t|$. The winding number  also equals the number of Majorana edge modes $W=\mathcal{M}$  and hence correctly represents the topological invariant in the static case.  Thus the  Majorana number jumps by $|\Delta{\cal M}| = 1$ across the TPT.

%%%%%%%%%%%%%%%%%%%%%%%%%%%%%%%%%
% SUBSECTION: RG-analysis for static Kitaev chain
%%%%%%%%%%%%%%%%%%%%%%%%%%%%%%%%%
\subsection{CRG analysis}
To extract the critical behavior around the TPT, we first obtain 
 the curvature function from  Eq.~(\ref{winding_number}), defined as
\begin{align}
F(k, {\bf M}) &\equiv \frac{\mathrm{d} \phi_k}{\mathrm{d}k}= \frac{a_{2,k} a_{3,k}' - a_{2,k}' a_{3,k}}{a_{3,k}^2 + a_{2,k}^2}
\nonumber \\
&= \frac{\mathrm{d}}{\mathrm{d} k} \arctan \left( \frac{t \cos k - \mu_0}{\Delta \sin k} \right) 
\nonumber \\
&= \frac{\Delta ( \mu_0 \cos k - t)}{(t \cos k - \mu_0)^2 + \Delta^2 \sin^2 k}.
\label{Scaling_function_static_Kitaev}
\end{align}
The winding number in Eq.~(\ref{winding_number})  is given by the  momentum-space integration of the  curvature function in Eq.~(\ref{winding_number_definition}).  Expanding around the high symmetry points,   $k_{0}=0$ or $\pi$, one can verify that the curvature function indeed manifests the Ornstein-Zernike form of Eq.~(\ref{Ornstein_Zernike_fit}). Some plots of $F(k, {\bf M})$ for different values of $\mu_0$ and $t=1$, $\Delta=0.7$ are shown in Fig.~\ref{plots-Berry-conn-static}. 

To implement the CRG procedure, we insert  Eq.~(\ref{Scaling_function_static_Kitaev}) into Eq.~(\ref{cond-RG-eqn}), and expand around the two HSPs $k_{0}=0$ and $k_{0}=\pi$ separately.  Choosing the
CRG parameter $M=\mu_{0}$ and using Eq.~(\ref{generic_RG_equation}), we obtain the following RG equation:
\begin{equation}
\frac{\mathrm{d} \mu_{0}}{\mathrm{d} l} = \pm t + \frac{\mu_{0}}{2} \mp \frac{\Delta^{2}}{t \mp \mu_{0}},
\label{RG_eq_static_mu}
\end{equation}
where the upper sign is for $k_{0}=0$ and the lower sign is for $k_{0}=\pi$.  This  RG flow  identifies the critical points $\mu_{0} =\pm t$ according to the rules in Eq.~\eqref{identifying_Mc_Mf}. The fixed points of  Eq.~\eqref{identifying_Mc_Mf} define ellipses centered at $(\mu_{0}, \Delta) =t (\mp\frac{1}{2}, 0)$:
\begin{equation}
\left( \frac{\mu_{0} \pm \frac{1}{2}}{\frac{3}{2}} \right)^2 + \left( \frac{\Delta}{\frac{3}{2\sqrt{2}}} \right)^2 = t^2.
\label{RG_eq_static_ellipses}
\end{equation}
with upper sign for $k_{0}=0$ and lower sign for $k_{0}=\pi$.

\begin{figure}[ht]
\centering
\includegraphics[width=\columnwidth]{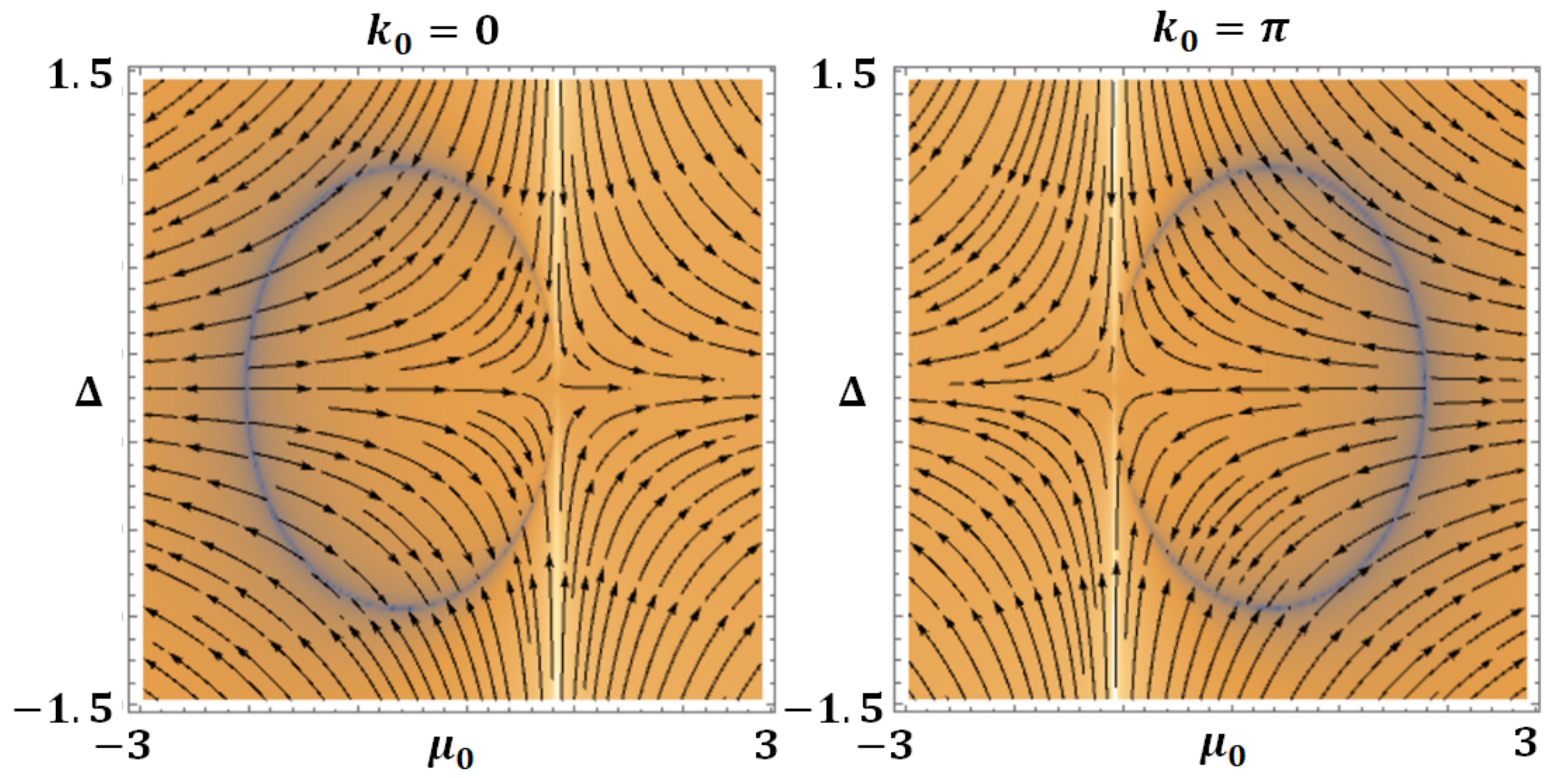}
\caption{RG flow of the static Kitaev chain described by Eqs.~(\ref{RG_eq_static_mu}) and (\ref{RG_eq_static_Delta}), using $k_{0}=0$ and $k_{0}=\pi$. The flow direction is indicated by the arrows, and the color scale indicates the flow rate in log scale. The yellow lines are the critical points $\mu_{0}=\pm t$ where the flow rate diverges, with
$t=1$ set to be the energy unit. The blue ellipses are the fixed points described by Eq.~(\ref{RG_eq_static_ellipses}) where the flow rate vanishes, which are stable in some 
regions and unstable in the other. }
\label{plots-RG-sol-static}
\end{figure}

The CRG procedure applied to the pairing gap $\Delta$ leads to
\begin{equation}
\frac{\mathrm{d} \Delta}{\mathrm{d} l} =  \mp \Delta \frac{\pm \Delta^2 + (t \pm \mu_{0}/2)(\mu_{0} \mp t)}{(t \mp \mu_{0})^2}.
\label{RG_eq_static_Delta}
\end{equation}
Eq.~(\ref{RG_eq_static_Delta})  represents the same set of critical points (vertical lines at $\mu_{0}=\pm t$) and fixed points (the two ellipses), as shown in Fig.~\ref{plots-RG-sol-static}. Note that the critical lines are independent of $\Delta$. 
%%%%%
%%%%%

%%%%%%%%%%%%%%%%%%%%%%%%%%%%%%%%%
% SUBSECTION: Majorana-Wannier state for static Kitaev chain
%%%%%%%%%%%%%%%%%%%%%%%%%%%%%%%%%
\subsection{Criticality - Majorana-Wannier state correlation function \label{sec:Majorana_Wannier_correlation}}

Next, we introduce a correlation function to quantify the proximity  to a TPT, according to Eq.~(\ref{lambdaR_Fourier_trans}).
Our intuition is based on previous investigations of other 1D non-superconducting systems, such as the Su-Schrieffer-Heeger (SSH) model, in which a Wannier state correlation function based on the Fourier transform of Berry connection is proposed~\citep{Chen:2017}.
For an analogous construction for the Kitaev chain, we  need to find an appropriate  gauge for the gauge-dependent Berry connection, such that we recover the curvature function in Eq.~(\ref{Scaling_function_static_Kitaev}) and the Ornstein-Zernike form cf. Eq.~(\ref{Ornstein_Zernike_fit}).

Firstly, we observe that the Dirac Hamiltonian in Eq.~\eqref{Dirac_model_static_Kitaev} has a filled-band eigenstate
\begin{eqnarray}
|u_{k-}\rangle=\frac{1}{\sqrt{2a_{k}(a_{k}+a_{3,k})}}
\left(
\begin{array}{c}
ia_{2,k} \\
a_{3,k}+a_{k}
\end{array}
\right)\;,
\label{wave_fn_uk}
\end{eqnarray}
where  $a_{k}=\sqrt{a_{2,k}^{2}+a_{3,k}^{2}}$. 
However, the corresponding Berry connection is trivial since  $A_{k}=\langle u_{k-}|i\partial_{k}|u_{k-}\rangle=0$.
To obtain a Berry connection of the form of Eq.~(\ref{Scaling_function_static_Kitaev}), we transform to the ``correct"  choice of  gauge:
\begin{eqnarray}
|\tilde{u}_{k-}\rangle=\frac{1}{\sqrt{2}a_{k}}
\left(
\begin{array}{c}
-a_{k} \\
a_{3,k}+ia_{2,k}
\end{array}
\right)=V_{k}|u_{k-}\rangle\;,
\label{Kitaev_chain_gauge_transformation}
\end{eqnarray}
such that  the Berry connection of this state  is  equal to (half of) the curvature function in Eq.~(\ref{Scaling_function_static_Kitaev}) 
\begin{eqnarray}
&&\tilde{A}_{k}=\langle \tilde{u}_{k-}|i\partial_{k}|\tilde{u}_{k-}\rangle
\nonumber \\
&&=\frac{a_{2,k}\partial_{k}a_{3,k}-a_{3,k}\partial_{k}a_{2,k}}{2a_{k}^{2}}
=\frac{F(k,{\bf M})}{2}
\nonumber \\
&&=\langle u_{k-}|i\partial_{k}|u_{k-}\rangle+\langle u_{k-}|\left(iV_{k}^{\dag}\partial_{k}V_{k}\right)|u_{k-}\rangle
\nonumber \\
&&=\langle u_{k-}|\left(iV_{k}^{\dag}\partial_{k}V_{k}\right)|u_{k-}\rangle\;.
\label{Berry_connection_gauge_choice}
\end{eqnarray}
Interestingly, $|\tilde{u}_{k-}\rangle$ is not an eigenstate of our Hamiltonian in Eq.~(\ref{Dirac_model_static_Kitaev}), but an eigenstate of 
\begin{eqnarray}
&&\tilde{h}(k)=Rh(k)R^{-1}=a_{3,k}\tau^{x}+a_{2,k}\tau^{y}\;,
\nonumber \\
&&R=e^{-i\tau_{y}\pi/4}=\frac{1}{\sqrt{2}}\left(
\begin{array}{cc}
1 & -1 \\
1 & 1
\end{array}
\right)\;,
\end{eqnarray}
\textit{i.e.}, rotating the particle-hole basis such that $a_{3,k}$ becomes the $a_{1,k}$ component. The eigenstate  basis of  $|\tilde{u}_{k-}\rangle$ is no longer the Nambu spinor $(f_{k},f_{-k}^{\dag})^{T}$, but the rotated spinor:
\begin{eqnarray}
R\left(
\begin{array}{c}
f_{k} \\
f_{-k}^{\dag}
\end{array}
\right)=\frac{1}{\sqrt{2}}\left(
\begin{array}{c}
f_{k}-f_{-k}^{\dag} \\
f_{k}+f_{-k}^{\dag}
\end{array}
\right)\;.
\label{Nambu_to_Majorana_rotation}
\end{eqnarray}
This new basis  has a nice physical interpretation as it is related to the momentum space operator of the real space Majorana fermions in Eq.~(\ref{Majorana_real_space}),
\begin{eqnarray}
f_{k}\pm f_{-k}^{\dag}=\sum_{n}e^{-ikr_{n}}\left(f_{n}\pm f_{n}^{\dag}\right)\;.
\end{eqnarray}
In summary, the gauge choice that leads to $|\tilde{u}_{k-}\rangle$ consists of (i) a rotation from the Nambu spinor to the Majorana basis, and (ii) a gauge choice  imposing an eigenstate specified by Eq.~(\ref{Kitaev_chain_gauge_transformation}).
This ensures that  the curvature function in Eq.~(\ref{Scaling_function_static_Kitaev}) and the Berry connection  shown in Eq.~(\ref{Berry_connection_gauge_choice}) are exactly equivalent.

The Majorana-Wannier states can now be defined in the $|\tilde{u}_{k-}\rangle$ basis
\begin{eqnarray}
&&|\tilde{u}_{k-}\rangle=\sum_{R}e^{-ik({\hat r}-R)}|R\rangle\;,
\nonumber \\
&&|R\rangle=\int dk e^{ik({\hat r}-R)}|\tilde{u}_{k-}\rangle\;.
\label{Wannier_state_definition}
\end{eqnarray}
This permits a direct transposition of the statement in the theory of charge polarization~\citep{King-Smith:1993,Resta:1994} to this Majorana problem: in this choice of gauge described in the previous paragraph, the winding number $W$ in Eq.~(\ref{winding_number_definition}) is equal to the charge polarization of the Majorana-Wannier state
\begin{eqnarray}
W=\int_{-\pi}^{\pi}\frac{{\rm d}k}{2\pi}\langle\tilde{u}_{k-}|i\partial_{k}|\tilde{u}_{k-}\rangle=\langle 0|{\hat r}|0\rangle\;,
\label{theory_of_charge_polarization}
\end{eqnarray}
where $|0\rangle$ denotes the Majorana-Wannier  function centered at the home cell $r=0$, and ${\hat r}$ is the position operator. 

Correspondingly,  the Fourier transform of the curvature function yields a Majorana-Wannier state correlation function~\citep{Chen:2017}  which in
conjunction with  the Ornstein-Zernike form of the curvature function in Eq.~(\ref{Ornstein_Zernike_fit}) yields
\begin{eqnarray}
\lambda_{R}&=&\int dk \langle \tilde{u}_{k-}|i\partial_{k}|\tilde{u}_{k-}\rangle e^{ikR}=\langle R|{\hat r}|0\rangle
\nonumber \\
&=&\int dr W^{\ast}(r-R)\; r\;W(r)\propto e^{-R/\xi_{k_0}}\;,
\label{Majorana_Wannier_correlation_static}
\end{eqnarray}
where $\xi_{k_0}$  is the correlation length at the relevant HSP. Note that the Majorana-Wannier state correlation function is nonzero in both topologically trivial and nontrivial phases, unlike the Majorana edge states that only appears in the topologically nontrivial phase. 
Near the critical point $\mu_{0}\rightarrow\pm t$, 
a straightforward expansion of Eq.~(\ref{Scaling_function_static_Kitaev}) around the HSP into the Ornstein-Zernike form of Eq.~\eqref{Ornstein_Zernike_fit} yields 
\begin{eqnarray}
&&\lim_{\mu_{0}\rightarrow\pm t}\xi_{k_0}=\left|\frac{\Delta}{\mu_{0}\mp t}\right|\;,
\nonumber \\
&&\lim_{\mu_{0}\rightarrow\pm t}F(k_{0},{\bf M})=\pm\frac{\Delta}{\mu_{0}\mp t}\;.
\label{static_Kitaev_xi_Fk0}
\end{eqnarray}
In the topologically nontrivial phase, the correlation length $\xi_{k_0}$ coincides with the localization length of the Majorana edge states, as proved explicitly in Appendix \ref{appendix:decay_length_Majorana_static}.
Comparing the above with  Eq.~(\ref{Fk0_xi_critical_exponent}), we immediately see that the critical exponents  defining the Kitaev chain TPT ($|\Delta{\cal M}|=1$) are  $\gamma=\nu=1$, compatible with the scaling law imposed by the conservation of the topological invariant within a phase. These critical exponents  are the same as those obtained for  other  TPTs in noninteracting static 1D Dirac models such as the SSH model \citep{Chen:2017}, indicating  that all these models belong to the same universality class.

%%%%%%%%%%%%%%%%%%%%%%%%
%%%%%%%%%%%%%%%%%%%%%%%%
% SECTION: DRIVEN KITAEV CHAIN
%%%%%%%%%%%%%%%%%%%%%%%%
%%%%%%%%%%%%%%%%%%%%%%%%
\section{Periodically driven Kitaev chain} 
\label{sec:periodically_driven_Kitaev}

%%%%%%%%%%%%%%%%%%%%%%%%
% SUBSECTION: Floquet-Majorana fermions
%%%%%%%%%%%%%%%%%%%%%%%%
\subsection{Floquet-Majorana fermions}

In this section, we extend the CRG approach to periodically driven 
Kitaev chain~\citep{Thakurathi,Molignini:2017}, which is known to host Floquet-Majorana modes.
The Hamiltonian describing the driven Kitaev chain is the same as in Eq.~(\ref{ham-fermionic-static}) plus a time modulation of the chemical potential:
\begin{align}
\mathcal{H}(t) \equiv \mathcal{H}_{0}[ \mu_0 \to \mu(t)].
\label{ham-fermionic}
\end{align}
For concreteness, we choose the driving to be a sequence of Dirac pulses with period $T$, $\mu(t) = \mu_0 + \mu_1 T \sum_{m \in \mathbb{Z}} \delta(t - m T)$.  The  momentum space equivalent of Eq.~(\ref{Dirac_model_static_Kitaev}) is obtained  using   $h_k(t) =  2\Delta \sin k \tau^y + 2(t \cos k - \mu(t)) \tau^z$.

To study the generation of Majorana modes,  we first obtain an effective Floquet Hamiltonian~\citep{Molignini:2017} describing the stroboscopic physics
\begin{equation}
h_{\text{eff},k} = i \log U_k(T,0),
\end{equation}
 from the time-ordered evolution operator
\begin{equation}
U_k(T,0) \equiv \mathcal{T} \left[ \exp \left( -i \int_0^T  h_k(t) \mathrm{d}t \right) \right]
\end{equation}
 since $U_k(NT, (N-1)T) = [U_k(T,0)]^N$ by virtue of the Floquet theorem. 
The eigenvalues of $h_{\text{eff}}$ define the quasienergies $\epsilon_{\alpha, k}$ of the Floquet-state solutions $\Psi_{\alpha}(k,t) = \exp(-i\epsilon_{\alpha,k} t) \Phi_{\alpha}(k,t)$, where $\Phi_{\alpha}(k,t) = \Phi_{\alpha}(k, t+ T)$~\citep{Haenggi}. 
The quasienergies are defined up to a multiple of $\frac{2\pi}{T}=\omega$ because of the $T$-periodicity of the Floquet modes  $\Phi_{\alpha}(k,t)$.
It is customary to restrict them to take values in a first ``{{Floquet-}}Brillouin zone'' of quasienergies $\epsilon_{\alpha} \in (-\omega/2, \omega/2)$. 

Akin to the static case, it is possible to identify  two kinds of zero-quasienergy Majorana modes in the Floquet spectrum, which are then termed Floquet-Majorana fermions (FMF's).
If  $\gamma_{\epsilon}^{\dagger}(t)$ denotes the creation operator of a Floquet mode $\Phi_{\epsilon}(t)$ associated to the quasienergy $\epsilon$ (we drop the index $\alpha$ for simplicity), then particle-hole symmetry implies $\gamma_{\epsilon}(t) = \gamma^{\dagger}_{-\epsilon}(t)$~\citep{Liu}. 
Hence, for $\epsilon=0$, we recover a zero-quasienergy Majorana mode as $\gamma_0(t) = \gamma_0^{\dagger}(t)$.
However, since quasienergies are defined within $(-\omega/2, \omega/2)$, the same situation can  occur for $\epsilon = \omega/2$ with $e^{-i \omega t /2} \gamma_{\omega/2} = \left(e^{-i \omega t /2} \gamma_{\omega/2} \right)^{\dagger}$~\citep{Liu}. 
Accordingly, the eigenvalues of $U(T,0) =\prod_k U_k(T,0)$ will then be either $e^{i 0 T} = 1$ or $e^{i \omega/2 T} = e^{i\pi} = -1$ and the corresponding  FMF's are labeled as $0$-FMF's or $\pi$-FMF's. 

FMF's have  characteristics  similar to their static counterparts: they have zero quasienergy (modulo $\omega/2$), have real wavefunctions (stemming from $\gamma = \gamma^{\dagger}$) and are localized near the ends of the chain~\citep{Thakurathi}. 
However, in contrast to the static case,  it is possible to generate  a hierarchy of FMF's by simply tuning the system's parameters over a wide range.
The system can hence be made topological even when the undriven phase has trivial topology 
and belongs to the  $\mathbb{Z}$ class, in contrast to the simpler $\mathbb{Z}_2$ categorization of the static system~\citep{Thakurathi, Molignini:2017}.

For the $\delta$-driving considered here,  the Floquet operator in $k$-space takes the form
\begin{equation}
U_k(T,0) = \left( \begin{array}{cc} A & B \\ -B & A^{*} \end{array} \right)
\end{equation}
where  $A=e^{2i \mu_1 T} \left( \cosh \omega(k) - i\frac{\beta(k)}{\omega(k)} \sinh \omega(k) \right)$, $B=-\frac{\alpha(k)}{\omega(k)} \sinh \omega(k)$, 
$\alpha(k) = 2T \Delta \sin k$, $\beta(k) = 2T(t \cos k  - \mu_0)$ and $\omega^2(k) = \beta^2(k) - \alpha^2(k)$.
Diagonalizing  $U_k(T,0)$,  we obtain the explicit form of the effective Hamiltonian as
\begin{equation}
h_{\mathrm{eff},k} = \frac{2\log \lambda^{-}}{\lambda^{+} - \lambda^{-}} \left[ \tilde{a}_{2,k} \tau^y + \tilde{a}_{3,k} \tau^z \right]
\label{heff_definition}
\end{equation}
where $\tilde{a}_{2,k} = B$ and $\tilde{a}_{3,k} = \Im[A]$ and $\lambda^{\pm}= \Re[A] \pm \sqrt{ \Re[A]^2 - (|A|^2  + |B|^2)}$ are the eigenvalues of $U_k(T,0)$.

Based on the similarity  of $h_{\text{eff},k}$ to the  static $h_k$, one could generalize the definition of the winding number introduced in Eq.~\eqref{winding_number} to the driven case. Nevertheless, it has been recently shown that this construction fails to correctly count the number of FMF's in certain driving regimes~\citep{Thakurathi, Rudner:2013}.
 An alternative method, obviating the calculation of micromotion, was proposed in Ref. \citep{Thakurathi} for the one dimensional case. 
It defines a finite line segment between the two points (modified according to our definition of the Hamiltonian)
\begin{eqnarray}
b_0(T, \mu_1)=\frac{2T}{\pi}\left(\mu_{0}-t+\mu_{1}\right)\;,
\nonumber \\
b_\pi(T, \mu_1)=\frac{2T}{\pi}\left(\mu_{0}+t+\mu_{1}\right)\;,
\end{eqnarray}
corresponding to the cases when the Floquet evolution operator $U_k(T,0) = \pm \mathds{1}$, which are realized at the HSPs $k=0$ and $k=\pi$ (hence the labelling). The topological invariant is then constructed from a non-trivial counting of the odd and even integers smaller/bigger than a certain threshold and encompassed by the segment, and can be written as $ \mathcal{M} = N_{0} + N_{\pi}$, where $N_{0(\pi)}$ counts the number of FMF's with Floquet eigenvalue $\pm 1$~\citep{Thakurathi}.

\begin{figure}[H]
\centering
\includegraphics[width=0.99\columnwidth]{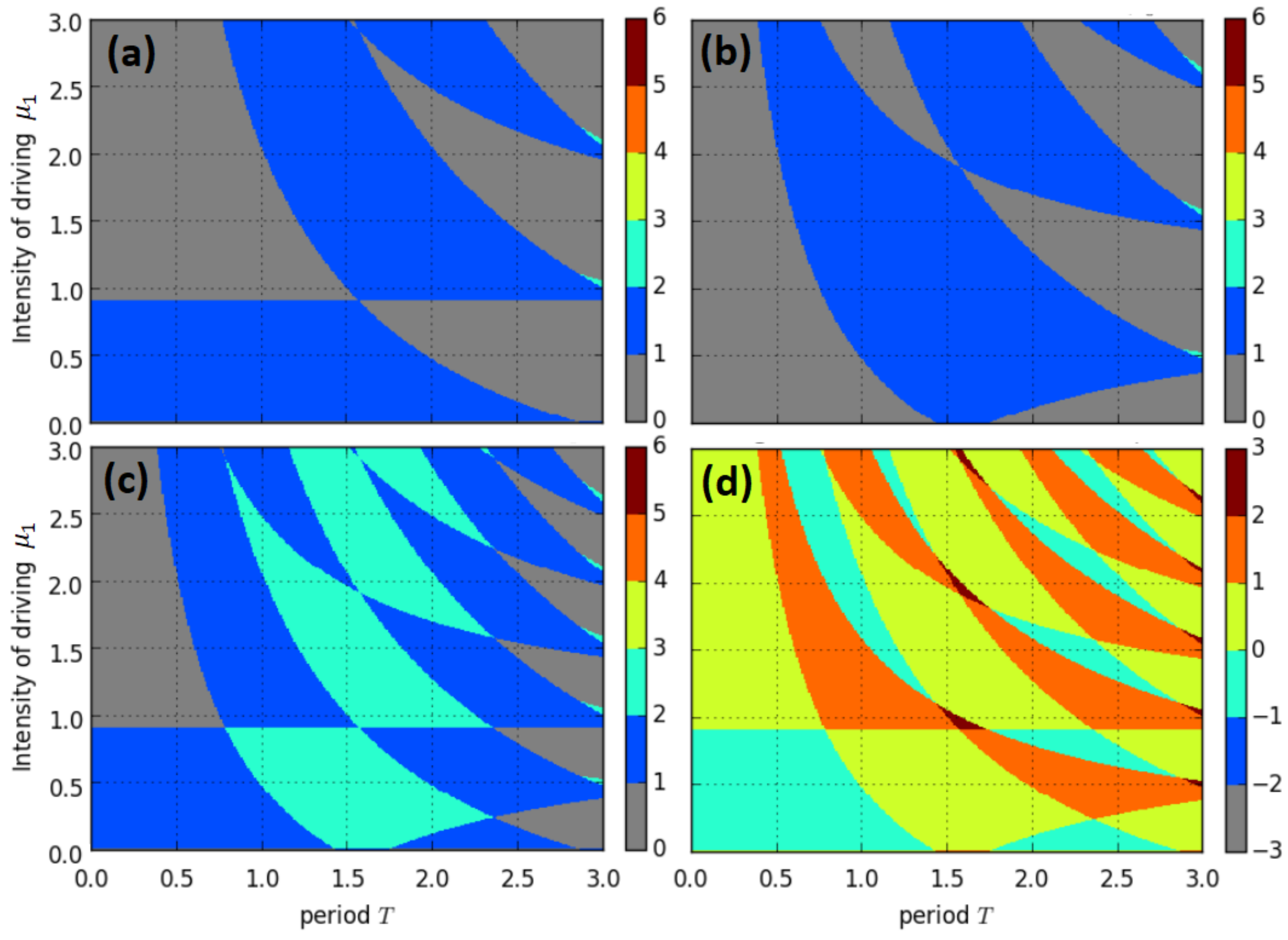}
\caption{Illustration of the number of FMF's of the driven Kitaev chain for $\mu_0=0.1$, plotted as a function of the driving parameters $T$ and $\mu_1$. The driving was performed starting from a static topological region. Note that the topological phase diagrams are independent of $\Delta$.
(a) The FMF's with Floquet eigenvalues $+1$. (b) The FMF's with Floquet eigenvalues $-1$. (c) The phase diagram of the system according to the total number of FMF's for each phase. (d) The winding number $W$ stemming from the Berry connection. One sees that the winding number $W$ does not fully coincide with the correct number of FMF's.}
\label{top-invariants1}
\end{figure}

By computing  both $N_0$ and $N_{\pi}$, we  map out the $(T,\mu_1)$-phase diagram of the inequivalent topological regions of the effective Floquet Hamiltonian.  Fig.~\ref{top-invariants1} depicts  such a phase diagram  and
clearly illustrates a mismatch with the one obtained from the conventional winding number calculations. Generally, the  static parameters $\Delta$ and $\mu_0$  strongly influence the  phase diagrams.  
 The  total number of FMF's   per edge $\mathcal{M} \in \mathbb{Z}$ as opposed to the static case where $\mathcal{M} \in \mathbb{Z}_2$ and can change by an integer across  the boundaries (see Fig.~\ref{top-invariants1}). 
Typically, from the phase diagram Fig.~\ref{top-invariants1}(c), we note that across the TPT $|\Delta{\cal M}| =1$.  Sometimes anomalous  transition regions exist at higher periods where $\Delta\mathcal{M}=2$. In what follows, we primarily focus on TPTs with $|\Delta{\cal M}| =1$ and discuss the anomalous lines with $|\Delta{\cal M}| = 2$ in section~\ref{sec:frozen-dynamics}.

%%%%%
%%%%%

\begin{figure}[ht]
\centering
\includegraphics[width=\columnwidth]{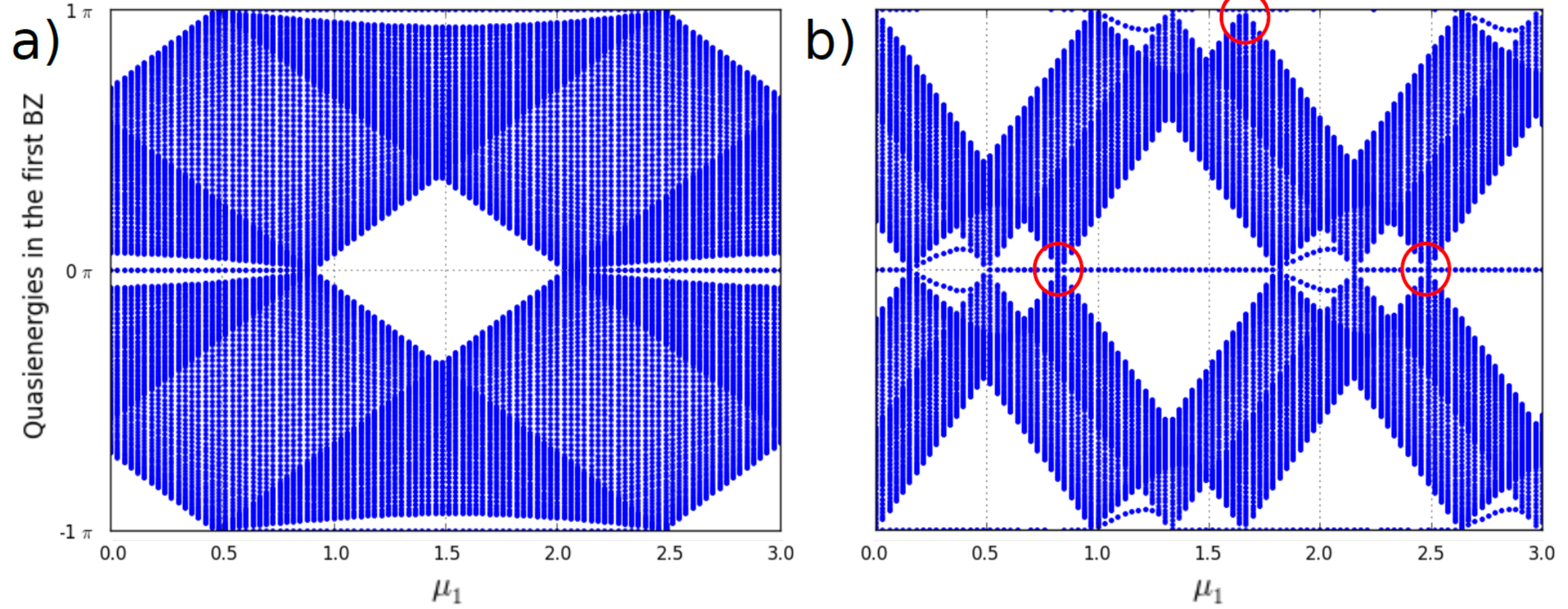}
\caption{Depiction of quasienergy spectra --- \textit{i.e.} eigenvalues of the Floquet operator --- as function of the driving intensity $\mu_1$ for an open chain of $N=100$ fermions. The other system's parameters are chosen as follows:
a) $\mu_0=0.1$, $\Delta=0.1$, $T=1.0$, 
b) $\mu_0=0.5$, $\Delta=0.9$, $T=1.9$.
The TPT's generating or removing $0$-FMF's ($\pi$-FMF's) are marked by gap closings at $0$ ($\pi$) with the corresponding appearance or disappearance of eigenvalues at $0$ ($\pi$). Note that there are instances of gap closings, marked by red circles, not associated with a change in the topological invariants. Those phase transitions appear in observables such as correlators and are detected by the CRG scheme.}
\label{quasienergy-spectra}
\end{figure}

To understand the TPTs, it is instructive to analyze the Floquet quasienergy dispersion, or equivalently the eigenvalues $e^{i \theta_k}$ of the Floquet operator $U_k$. They can be compactly written as~\citep{Molignini:2017}
\begin{align}
\cos \theta_k &= \cos(2\mu_1 T) \cos(T E_{k}) +  \nonumber \\
& \quad +  \sin(2\mu_1 T) \frac{2(t \cos k - \mu_0)}{E_{k}} \sin(T E_{k})
\label{quasienergy_delta_func_driving}
\end{align}
with the static energy dispersion
\begin{equation}
E_{k} = 2 \sqrt{(t \cos(k) - \mu_0)^2 + \Delta^2 \sin^2(k)}.
\end{equation}
In accordance with  the bulk-edge correspondence~\citep{Kitaev-table}, TPT's should be signalled by a closing of the gap  in the quasienergy spectrum.  We note that instances of TPTs not associated with gap closing have been discovered in systems where the symmetry of the Hamiltonian changes across the topological phase boundary~\citep{Ezawa:2013}.  
Our results for 
the quasienergy spectra  for long but  finite Kitaev chains are  shown in Fig.~\ref{quasienergy-spectra}.  Note that there are gap closings at  $0$ or $\pi$ quasienergies, reflecting the creation or annihilation of $0,\pi$-Majorana modes. 

\begin{figure}[h]
\centering
\includegraphics[width=0.99\columnwidth]{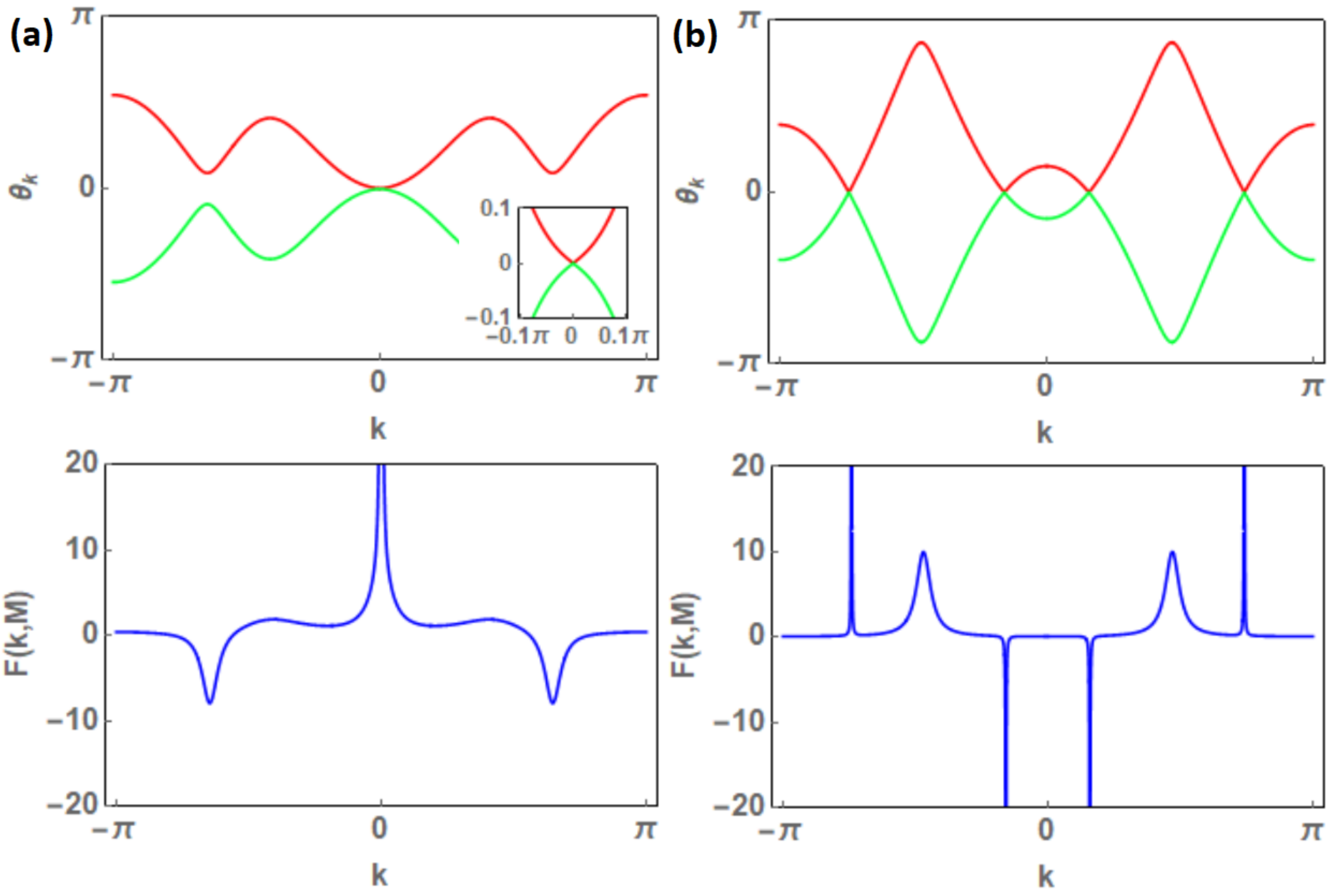}
\caption{ (Top) Gap-closing manifests in the quasienergy dispersion $\theta_k$, and (bottom) the corresponding divergence of the curvature function $F(k,{\bf M})$ in the periodically driven Kitaev chain: (a) At $\Delta=0.5$, $\mu_0=0.1$, $\mu_1=0.9$, $T=2.0$, which is the critical point of gap-closing at $k_{0}=0$ and creating a $0$-FMF. The inset of the quasienergy plot shows that the gap-closing at $k_{0}=0$ is in fact linear at low energy, although it looks quadratic at larger scale. (b) At $\Delta=0.1$, $\mu_0=0.1$, $\mu_1=0.78$, $T=2.0$, where the gap closes at non-HSPs (see also Fig.~\ref{figure:backfolding}). }
\label{gap-closings-quasienergy}
\end{figure}

Additionally, a systematic analysis of the bulk quasienergy dispersion $\theta_k$  for different values of the static parameters $\Delta$ and $\mu_0$, reveals gap closures  with linear dispersions
around  $\theta_{k}=0$  and the zone edge ($\theta_{k}=\pi$) whenever a new $0$-FMF ($\pi$-FMF) is generated or destroyed. These gap closures related to TPTs, specifically appear 
at HSP  $k=0$ and $k=\pi$ in analogy with  the static case.
This  behaviour, shown in Fig.~\ref{gap-closings-quasienergy}(a),  is in agreement with  expectations for  the universality class $|\Delta{\cal M}|=1$.

%%%%%
%%%%%

%%%%%
%%%%%

Note that gap closures can also occur in  $\theta_k$ at \textit{non}-HSP (see Fig.~\ref{gap-closings-quasienergy}(b)).  
These features at non-HSP are  visible in the quasienergy spectra  but are  \textit{not} systematically associated with a change in the $N_0 + N_{\pi}$ (see Fig.~\ref{top-invariants1}d)).
We will show in the next section that the CRG scheme is  capable of capturing the physics of  both topological and nontopological gap closures at  HSP and non HSP.

%%%%%%%%%%%%%%%%%%%%%%%%%
% SUBSECTION: RG flow for driven Kitaev chain
%%%%%%%%%%%%%%%%%%%%%%%%%
\subsection{RG flow for the Floquet effective Hamiltonian of the driven Kitaev chain}

In the previous sections, we saw that driving can indeed generate a hierarchy of Floquet-Majorana modes $\mathcal M$ with TPTs between zones with differing $\mathcal M$.
We now demonstrate that this complex phase diagram --- cf. Fig.~\ref{top-invariants1} --- can be obtained by the CRG procedure outlined in Sec.~\ref{sec:static_kitaev_chain}. As in the static case, the effective Floquet Hamiltonian $h_{\text{eff}}$ in Eq.~\eqref{heff_definition} defines an angle function $\phi_k$ that represents the angle that $h_{\text{eff}}$ spans in the $yz$-plane.
The curvature function $F(k,{\bf M}) \equiv \frac{\mathrm{d}\phi_k}{\mathrm{d} k}$ is calculated from the angle function as
\begin{widetext}
\begin{align}
F(k,{\bf M}) &= \frac{\mathrm{d}}{\mathrm{d}k} \arctan \left( \frac{\tilde{a}_{3,k}}{\tilde{a}_{2,k}} \right) = \frac{\mathrm{d}}{\mathrm{d}k}  \arctan \left[  \frac{ \cos(2\mu_1T) (t \cos (k) - \mu_0) \sin(T E_{k}) - \frac{\sin(2\mu_1 T)}{2} \cos(T E_{k}) E_{k}}{\Delta \sin (k) \sin(T E_{k})} \right].
\label{Scaling_function_deriven_Kitaev}
\end{align}
\end{widetext}
Sample curvature functions are shown in the  { the bottom panels of} Fig.~\ref{gap-closings-quasienergy}.

%%%%%
\begin{figure}[h]
\centering
\includegraphics[width=0.99\columnwidth]{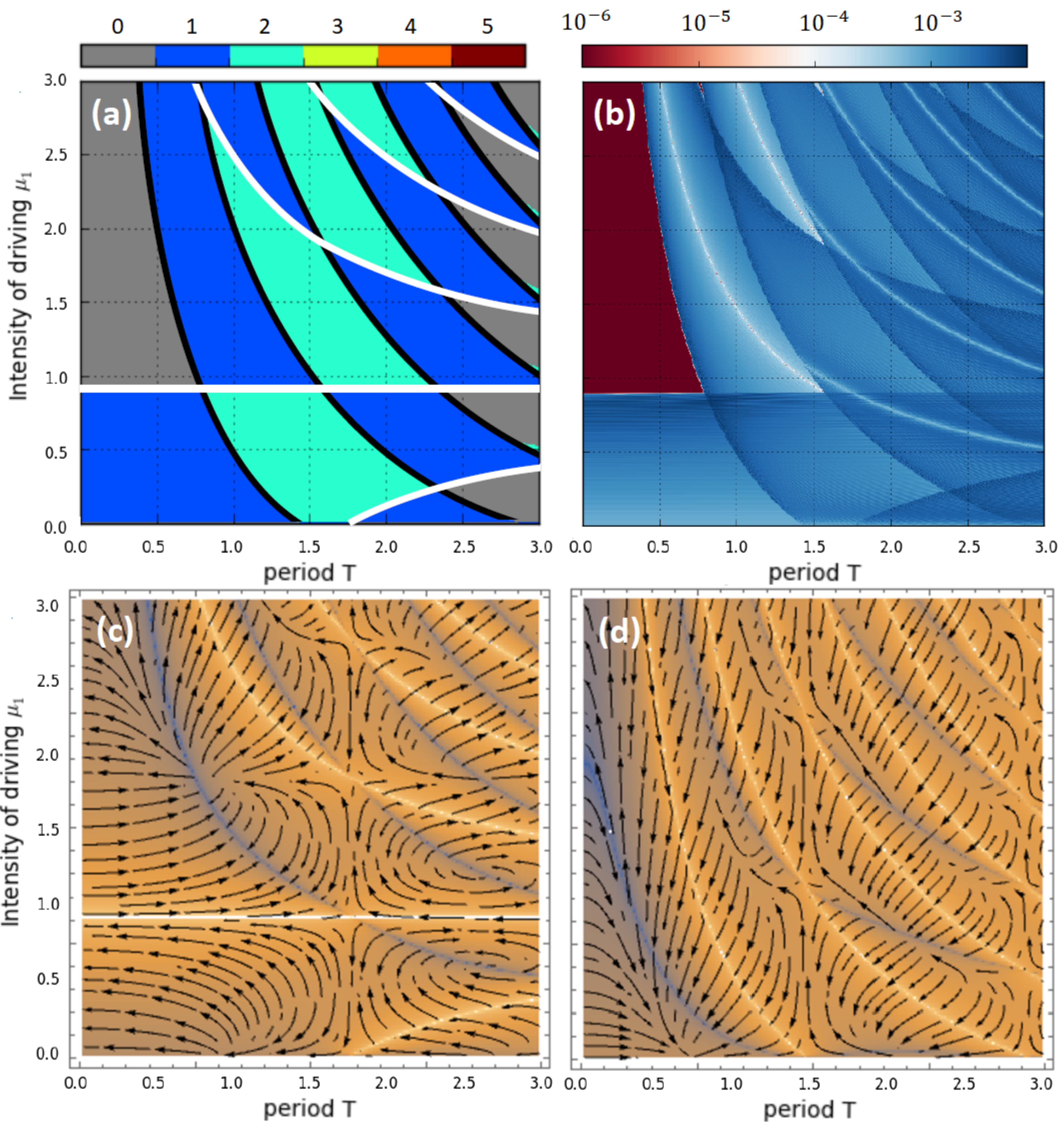}
\caption{ (a) The topological phase boundaries in the ${\bf M}=(T,\mu_1)$ parameter space for the periodically driven Kitaev chain at $\mu_{0}=0.1$. White lines signal the creation of a $0$-FMF and black lines the creation of a $\pi$-FMF. (b) The magnitude of the residual correlator $C_{\text{res}}$ in the same parameter space, taken from Ref.~\citep{Molignini:2017}. (c) The RG flow obtained from $k_{0}=0$. The color codes are log of the numerator $\log \left[\partial_{k}^{2}F(k,{\bf M})|_{k=0}\right]$ in Eq.~(\ref{generic_RG_equation}), with orange the high value and blue the low value. The bright lines (critical points of $0$-FMF) coincide with the white lines in (a), and blue lines (stable or unstable fixed points) are close to the white lines in (b) (minimum of $C_{\text{res}}$). (d) The RG flow obtained from $k_{0}=\pi$, whose bright lines (critical points of $\pi$-FMF) correspond to the black lines in (a)}
\label{figure:RG-comparison}
\end{figure}
%%%%%

In the periodically driven case, the parameters that define the CRG flow are ${\bf M} = (T, \mu_1)$, \textit{i.e.} the  period and intensity of the driving.
As for the static case, we use Eq.~\eqref{generic_RG_equation} to obtain the flow equations for the system in the ${\bf M}=(T,\mu_{1})$ parameter space. However, since the analytical expressions of the derivatives of the curvature function are cumbersome, we numerically evaluate them on a discrete mesh of points:
\begin{eqnarray}
\frac{dM_{i}}{dl}=\frac{\Delta M_{i}}{\Delta k^{2}}\frac{F(k_{0}+\Delta k,{\bf M})-F(k_{0},{\bf M})}{F(k_{0},{\bf M}+\Delta {\bf M}_{i})-F(k_{0},{\bf M})}\;,
\label{RG_eq_numerical}
\end{eqnarray}
where $\Delta k$, $\Delta{\bf M}=(\Delta T,\Delta\mu_{1})$ are grid spacings. The advantage of Eq.~(\ref{RG_eq_numerical}) is that at each mesh point of the ${\bf M}=(T,\mu_{1})$ parameter space, one only requires to calculate two points in the momentum space $k_{0}$ and $k_{0}+\Delta k$ without explicitly performing the integration of winding number in Eq.~(\ref{winding_number_definition}), rendering a very convenient numerical tool to identify TPT.
The resulting RG flows for both HSP $k=0$, $k=\pi$ are shown in Figs.~\ref{figure:RG-comparison}(c) and (d) for $\mu_0=0.1$. 
Using the criteria outlined in Eq.~\eqref{identifying_Mc_Mf}, we distinguish the set of critical and fixed points as the bright lines of maximal flow and the dark lines of zero flow, respectively.

\subsubsection{Critical Lines}
Comparing with the phase diagram obtained Fig.~\ref{top-invariants1}(c), we see that the critical lines of the CRG method correctly capture the phase boundaries where $\Delta\mathcal{M}=1$.
The CRG scheme is also able to track which type of FMFs are created or annihilated at the critical boundaries.
A direct comparison of the flow diagrams with Fig.~\ref{figure:RG-comparison}(a) and (b) reveals that the phase boundaries where the number of 0($\pi)$-FMF's $N_{0}$($N_\pi$) changes are delineated by the critical lines of the CRG flow around the HSP $k_0=0(\pi)$, respectively.
This  correlation holds true also for all ranges of the parameters $\mu_0,\mu_1$ and $T$.

\begin{figure}[ht]
\centering
\includegraphics[width=\columnwidth]{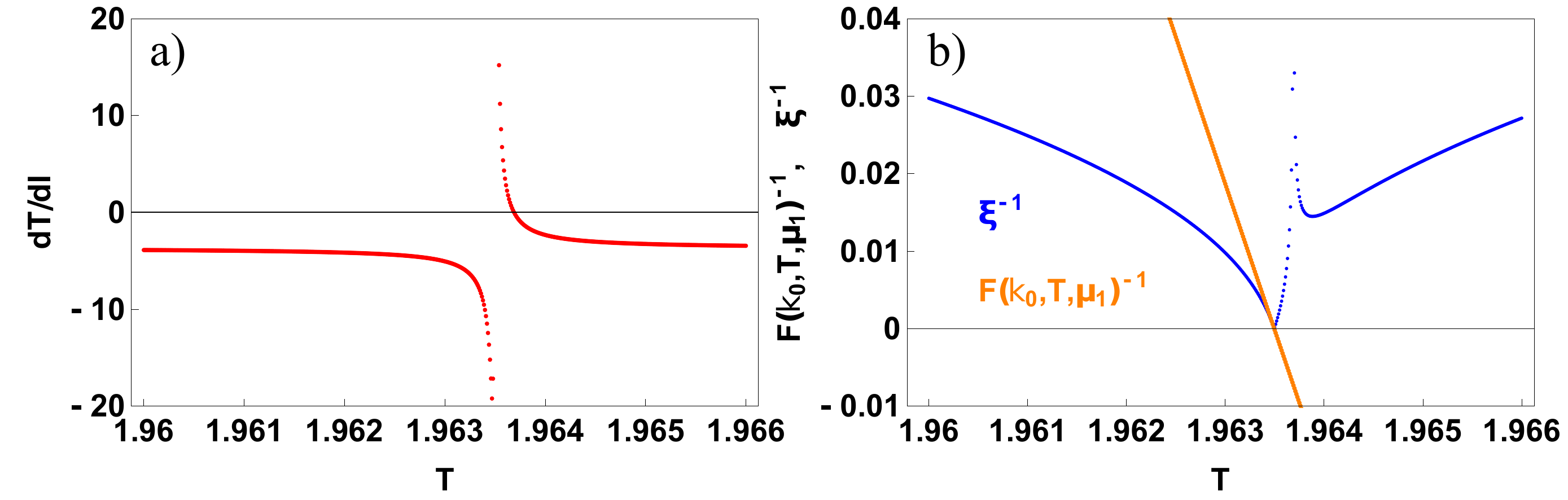}
\caption{ (a) The CRG flow along $T$-direction at fixed $\mu_{1}=0.1$ close to the critical point ${\bf M}_{c}=(T_{c},\mu_{1c})\approx(1.9635,0.1)$. The flow has been obtained from $k_0 =0$ by numerically evaluating the derivatives with a very find grid $\Delta k=0.0001$ and $\Delta T=0.00001$. One sees that the critical point $\mu_{1c}\approx 1.9635$ at which $dT/dl$ diverges and the fixed point $\mu_{1f}\approx 1.9637$ are extremely close. (b) The inverse of the Wannier state correlation length $\xi_{k_0}^{-1}$ and that of the curvature function at HSP $F(k_{0},T,\mu_{1})^{-1}$, both vanish linearly as $T\rightarrow T_{c}$, indicating their critical exponents $\gamma=\nu=1$.}
\label{fig:driven_RGflow_detail}
\end{figure}

Analytical expressions for the critical flow lines where $N_0$ or $N_{\pi}$ changes by one can also be obtained by analyzing 
the divergences of $\lim_{k \to 0, \pi} F(k, T, \mu_1)$ as a function of $T, \mu_1$.  The TPTs occur at the boundaries defined by the simple equations:
\begin{align}\label{anal}
0-\text{FMF}: \quad \mu_1(T) &= \frac{m_0 \pi}{2T} + (t - \mu_0), \quad m_0 \in \mathbb{Z} \nonumber \\
\pi-\text{FMF}: \quad \mu_1(T) &= \frac{m_{\pi} \pi}{2T} - (t + \mu_0), \quad m_{\pi} \in \mathbb{Z},
\end{align}
which agrees with the boundaries delineated by $\mathcal M$ in  Fig.~\ref{figure:RG-comparison}(a) and the residual correlator in Fig.~\ref{figure:RG-comparison}(a)-(b).
We note that  there are fixed lines ${\bf M}_{f}$   in close proximity to some dominant critical lines ${\bf M}_{c}$( not visible in the flow)  and hence subsumed by the criticality. 
To distinguish ${\bf M}_{c}$ from ${\bf M}_{f}$, very fine spacings of $\Delta k$, $\Delta T$ and $\Delta\mu_{1}$ must be used, as demonstrated in Fig.~\ref{fig:driven_RGflow_detail} (a).
A striking feature of \eqref{anal} is that as in the static case,   the  critical lines and hence the topological phase diagram in the Floquet case are  independent of  the anisotropy/p-wave parameter $\Delta$. This is a non-trivial
prediction which can easily be verified  using the quasienergy spectra.

Signatures of these transitions in a driven-dissipative setup were recently studied in ~\citep{Molignini:2017, Prosen2011}.
It was  shown  that the  weighted sum of the Majorana correlator $C_{ij}(t) =\left<\omega_i \omega_j\right> - \delta_{ij}$ was capable of delineating the boundaries between
different topological phases.
This sum, termed the \textit{residual correlator}, is defined as $C_{\text{res}} \propto \sum_{|j-k| \ge N/2} |C_{jk}|$~\citep{Prosen2011, Molignini:2017} and
 effectively filters out short-range correlations and offers a measure of long-range correlations.
A typical plot of $C_{\text{res}}$ from Ref.~\citep{Molignini:2017} is shown in Fig.~\ref{figure:RG-comparison}.
Clearly, the residual correlator is able to precisely track the critical topological phase boundaries.
Note that lines of pronounced low correlation are seen in  $C_{\text{res}}$ which are unrelated to the critical lines.
We will show in Sec.~\ref{sec:frozen-dynamics} that these low correlation lines are related to fixed lines of the CRG and frozen dynamics.

%%%%%
%%%%%

%%%%%%%%%%%%%%%%%%%%%%%%%%%%%%%%%
% SUBSECTION: Majorana-Wannier state for driven chain
%%%%%%%%%%%%%%%%%%%%%%%%%%%%%%%%%

\subsubsection{Wannier state correlation functions and critical exponents}

To characterize the TPTs  for the Floquet chain, we derive the equivalent  Majorana-Wannier state correlation function. We  follow the same procedure outlined for the static case in Sec.~\ref{sec:Majorana_Wannier_correlation}.
Let $|u_{k-}\rangle$ denote the lowest eigenstate of $h_{\text{eff},k}$. It has  the form of Eq.~(\ref{wave_fn_uk}) and a vanishing Berry connection $A_{k}=\langle u_{k-}|i\partial_{k}|u_{k-}\rangle=0$.
We use the gauge transformation Eq.~(\ref{Kitaev_chain_gauge_transformation}) to obtain
$\tilde{h}_{\text{eff},k}=Rh_{\text{eff},k}R^{-1}=\tilde{a}_{3,k}\tau^{x}+\tilde{a}_{2,k}\tau^{y}$.
The eigenstate of $\tilde{h}_{\text{eff},k}$, $|\tilde{u}_{k-}\rangle$ is  again expressible in the basis of Floquet-Majorana fermions as in Eq.~(\ref{Nambu_to_Majorana_rotation}). 
The nonvanishing  Berry connection of Eq.~(\ref{Scaling_function_deriven_Kitaev})  for the driven case is now given by $\tilde{A}_{k}=\langle \tilde{u}_{k-}|i\partial_{k}|\tilde{u}_{k-}\rangle=F(k,{\bf M})/2$.
The Majorana-Wannier state $|R\rangle$   and the corresponding  Majorana-Wannier state correlation function can then be defined using  Eqs.~(\ref{Wannier_state_definition}) and ~(\ref{Majorana_Wannier_correlation_static}).
$|R\rangle$ is now a time-independent \textit{stroboscopic} function encoding the physics in $h_{\text{eff}}$.

To obtain the critical exponents of the Floquet TPTs, we  compute the Majorana-Wannier state correlation length.  We find that
\begin{eqnarray}
\xi_{k_0} \approx\left|\frac{(\partial_{k}\tilde{a}_{2,k})_{k_{0}}}{\tilde{a}_{3,k_{0}}}\right|
=\left|F(k_{0},{\bf M})\right|\;,
\label{driven_xi_Fk0_bk}
\end{eqnarray}
implying that the critical exponents of the TPT $\gamma=\nu$.
The detailed calculations are performed in Appendix \ref{appendix:decay_length_Majorana_driven}.
Similar to the static case,  there are two correlation lengths $\xi_0$ and $\xi_{\pi}$ depending on which HSP is considered. At the TPTs, only the $\xi_{k_0}$ associated with the HSP $k_0$ at which the gap closes in the quasienergy spectrum  will diverge.
Close to the TPT  driven by a control parameter $M_i$ as calculated in Appendix \ref{appendix:decay_length_Majorana_driven},
\begin{eqnarray}
\xi_{k_0} \propto \frac1 {\vert M_i -M_{ic}\vert}
\label{driven_xi_Mi}
\end{eqnarray} 
where $M_{ic}$ is the critical value of the parameter.
This demonstrates that the critical exponents of the Floquet TPT are $\nu=\gamma=1$. 
This result is further bolstered by  the numerical extraction of the exponents from the 
 Ornstein-Zernike fit to the curvature function $F(k,{\bf M})$ in Eq.~(\ref{Ornstein_Zernike_fit}). 
From Fig.~\ref{fig:driven_RGflow_detail} (b), we see  that  $\xi_{k_0}^{-1}$ is linear in   $M_i -M_{ic}$ only in a very narrow range near ${\bf M}_{c}$, indicating that the critical region is in general very small.

In Appendix \ref{appendix:decay_length_Majorana_driven}, we show that the Floquet Majorana edge state {\it being created at the TPTs} has a decay length that coincides with the Majorana-Wannier state correlation length.
Across each critical boundary in Fig.~\ref{figure:RG-comparison} {\it only one Majorana edge state is created}, \textit{i.e.} $|\Delta{\cal M}|=1$, and this state determines the critical behavior.
The change in  Majorana number   $|\Delta{\cal M}|=1$  in conjunction with the  critical exponents  $\nu=\gamma=1$ means that the driven system still belongs to the same universality class as the static Majorana chain, despite the intrinsically more complex  phase diagram of the Floquet Majorana chain  cf. Fig.~\ref{figure:RG-comparison}.

%%%%%%%%%%%%%%%%%%%%%%%%%%%%%%%%%
% SUBSECTION: LCFLs and frozen dynamics
%%%%%%%%%%%%%%%%%%%%%%%%%%%%%%%%%
\subsubsection{Fixed lines and frozen dynamics}
\label{sec:frozen-dynamics}

The fixed lines of the CRG resolve the puzzle of low correlations in $C_{\text{res}}$ discussed earlier.  Specifically,  comparing Figs.~\ref{figure:RG-comparison}(b), (c) and (d) we see  that the combined  fixed lines of the flow ${\bf M}_{f}$  for   $k_0=0$ and $k_0=\pi$ precisely encompass  the regions of reduced correlations  in $C_{\text{res}}$. This correspondence is sound, because the fixed lines represent points where the Majorana-Wannier state correlation length vanishes (see next section). 

Insight into the nature of the fixed lines (FL) can be obtained by examining the behavior of the Berry connection as one traverses these lines.
We find that  the  Berry connection shows divergences  at non HSPs which  then flip sign  across the  FL if $\mathcal{M} > 1$. 
When $\cal{M} \le 1$ no divergence occurs like along the fixed lines in the static case.

%%%%%
\begin{figure}[ht]
\centering
\includegraphics[width=0.99\columnwidth]{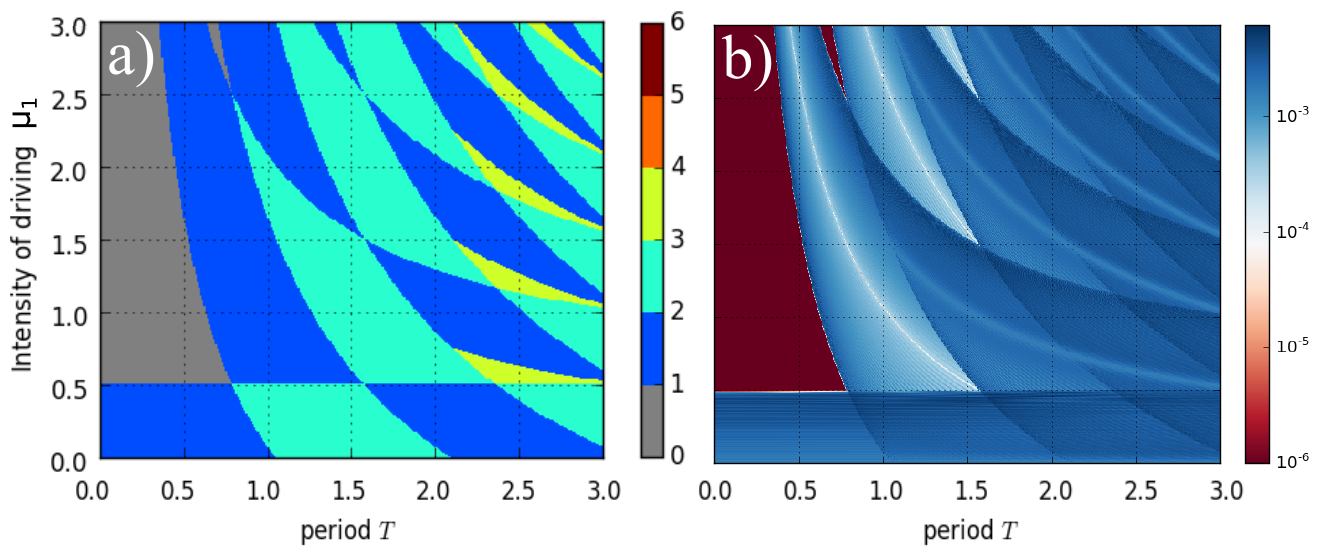}
\caption{Comparison between the topological invariant $\mathcal{M}=N_{0} + N_{\pi}$ and the residual correlator $C_{res}$ for the $(T,\mu_1)$-phase diagram of the driven Kitaev chain with $\Delta=0.1$, $\mu_0=0.5$. The topological invariant seems to indicate additional phase boundaries at higher periods, where the number of FMF's jumps by two. These additional phase boundaries coincide with the fixed lines of the CRG flow and the lines of low correlation in $C_{res}$.}
\label{top-invariants2}
\end{figure}
%%%%% 

%%%%%
\begin{figure}[ht]
\centering
\includegraphics[width=\columnwidth]{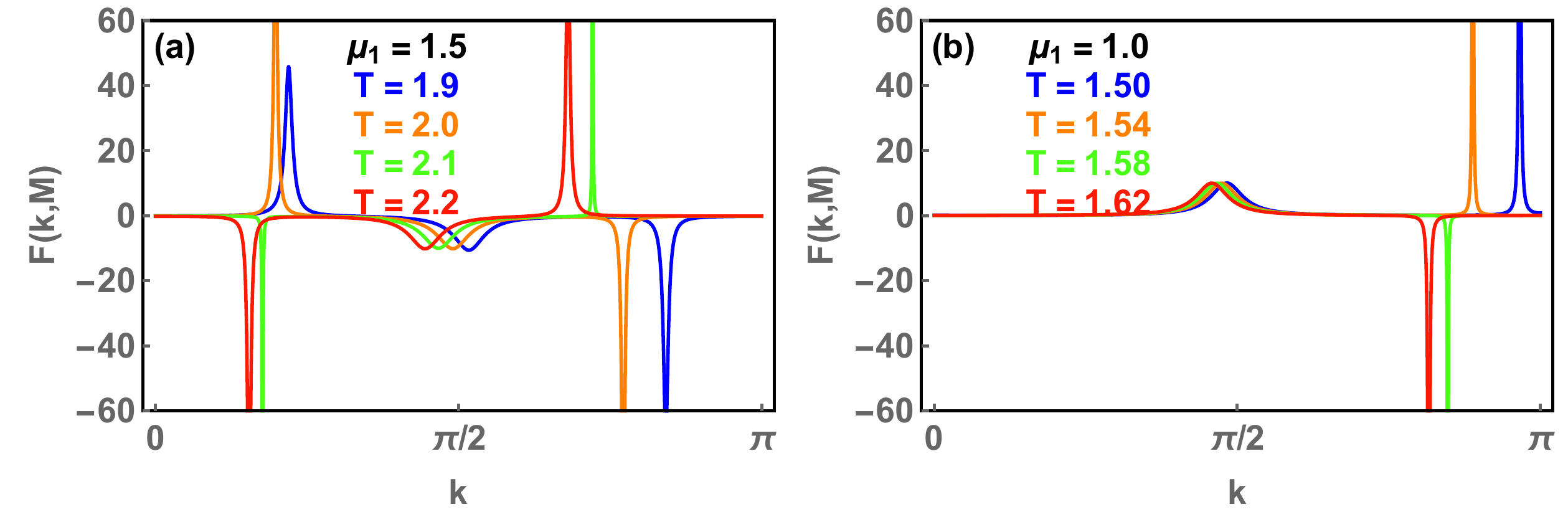}
\caption{Plot of the Berry connection $F(k, {\bf M})$ across the FL for $\Delta=0.1$, $\mu_0=0.1$, and at a) $\mu_1=1.5$ and b) $\mu_1=1.0$. Panel a) shows the Berry connection for a region with one FMF (see also Fig.~\ref{top-invariants1}). The same number of positive and negative peaks flips sign. Hence, the topological index $W=-1$ is unchanged during this transition. Panel b) shows instead a region with two FMF's, while the topological index $W$ changes from $0$ to $+2$. Note that the smaller peaks centered around $k \approx 1.4$ have the same weight (area under the peak) as the diverging peaks that flip sign. 
}
\label{figure:peak-flip-DeltaW}
\end{figure}
%%%%%

\begin{figure}[ht]
\centering
\includegraphics[width=\columnwidth]{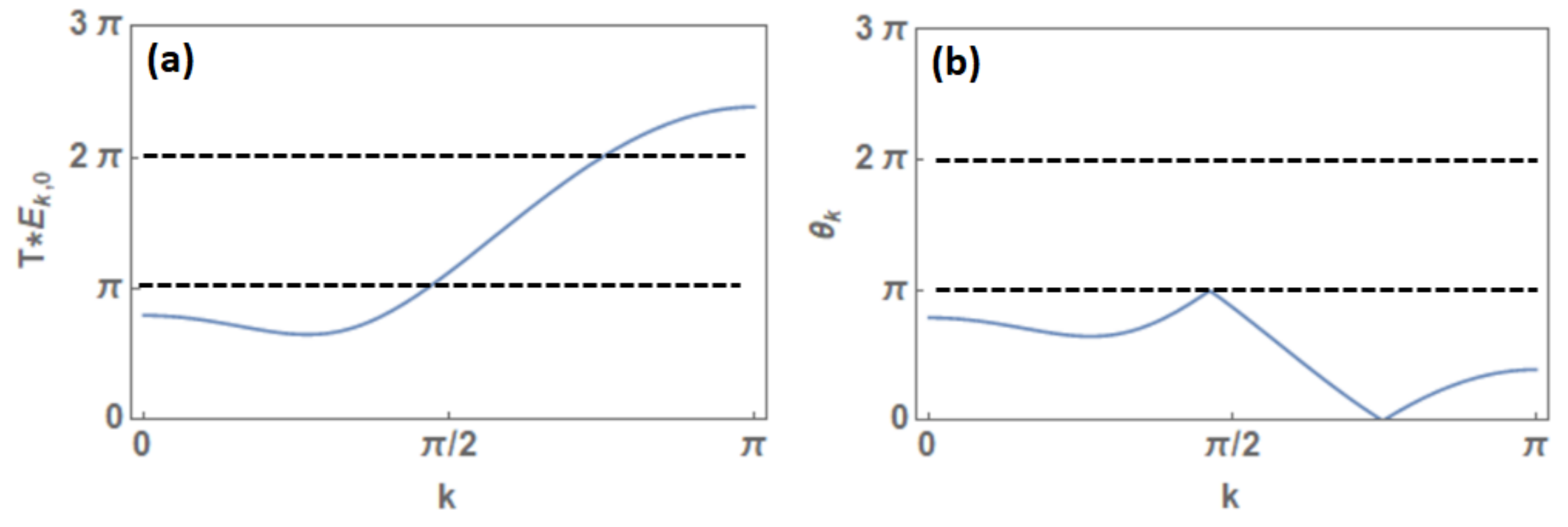}
\caption{Backfolding of the upper band of (a) the static energy dispersion into (b) the first Floquet-Brillouin zone corresponding to the LCFLs where the dynamics is frozen. This backfolding procedure induces apparent gap closings at non-HSPs. The parameters are $\Delta=0.1$, $\mu_0=0.1$, $\mu_1=1.26$ and $T=2.5$.}
\label{figure:backfolding}
\end{figure}

Remarkably, the CRG approach based on the expansion around HSPs captures very well the divergences at non-HSPs. 
This can be attributed to the conservation of the winding number $W$ within a phase, such that when the divergence at non-HSP occurs, the curvature function at HSP will be affected.
We now discuss the possible physical mechanism leading to the FL and how to obtain an analytical forms for their equations.
A closer inspection of the quasienergy dispersion \eqref{quasienergy_delta_func_driving} reveals that, for $\mu_1(T) = \frac{\pi m}{2T}$ with $m \in \mathbb{Z}$, the second summand vanishes because of the sine being zero. 
We are thus left with~\citep{Prosen2011}
\begin{equation}\label{theta}
\theta_k= \begin{cases} 
T E_{k}, \quad &m \in 2\mathbb{Z} \\
\pi - T E_{k}, \quad &m \in 2\mathbb{Z}+1.
\end{cases}
\end{equation}
The lines described by the form of $\mu_1(T)$ above correspond to the location of the FL.
Consequently, along FLs the driving is momentarily frozen and the quasienergy dispersion is the static energy dispersion $E_{k}$, where the relevant topology can host at most one Majorana mode per edge. 
This also explains why  gap closings appear at non-HSPs: along the FL, for $k$ values where $\vert E_{k} \vert  > \pi$, the constraint $-\pi\le \theta_k \le \pi$ requires that $|E_{k}|$ to be folded back to the first Floquet-Brillouin zone, as illustrated in Fig.~\ref{figure:backfolding}. The nodes in the folding, which occur at non-HSPs, reminisce gapless points.  It remains to be seen if the peculiar behavior of the FL associated with frozen dynamics is a feature of the type of driving applied, or if it can be extended to other driving protocols such a two- or multistep-driving.

A natural question is whether the FLs signal additional TPTs.
Comparing Figs.~\ref{top-invariants2}(a) and \ref{top-invariants2}(b), we clearly see that the FLs are not systematically associated with a change in the number of FMF's given by $\mathcal{M} = N_{0} + N_{\pi}$.
However,  there exist  regions, where the FLs  signal a change in FMFs with $\Delta\mathcal{M}=2$. An example is  the yellow regions  in Fig.~\ref{top-invariants2}(a)  where $\mathcal{M}=3$ jumps to $\mathcal{M}=1$ .
Such $|\Delta{\cal M}|=2$  jumps are also seen for other parameters.
We have verified that these jumps are indeed systematically associated with gap closings at $0$ or $\pm \pi$ in the Floquet quasienergy spectrum for both open and closed systems with $N=2000$. The corresponding wave functions are real and show edge localization for open chains (for details see Appendix \ref{appendix:figures_2Maj_jumps}).
These gap closures indicate that sometimes FLs  signal transitions where $\Delta\mathcal{M} \neq \pm 1$   (see Fig.~\ref{gap-closings-quasienergy} and \ref{quasienergy-spectra}).   
It follows that these do not belong to the universality class of the TPTs where $\Delta\mathcal{M} =\pm 1$. 

These intriguing features merit further study to verify if indeed a driven system can host multiple
universality classes or if they are in anyway related to the anomalous topological phases discussed in driven two dimensional systems\citep{Rudner:2013}, where new FMF's modes are indeed created but the \emph{difference} of the FMF's $|N_0 - N_{\pi}|$ remains unchanged (\textit{e.g.} $N_0=1, N_{\pi}=0 \to N_0 =1, N_{\pi}=2$).

%%%%%
%%%%%

%%%%%%%%%%%%%%%%%%%%%%%%%%%%%%%%%
% SECTION: Conclusions
%%%%%%%%%%%%%%%%%%%%%%%%%%%%%%%%%
\section{Conclusions}
\label{sec:conclusions}

In summary, we have applied the CRG approach to study TPTs in  static and periodically driven Kitaev chains.
The {method}, though insensitive to micromotion, provides a simple and efficient way to obtain the full topological phase diagram of the driven system.
Comparing our results with quasienergy spectra calculations, as well as exact Majorana correlation functions,  we find that the CRG scheme correctly captures topological phase boundaries.  The critical points of the CRG flow correspond to TPTs where the number of localized edge Majorana modes changes by one. 
Extending the notions of charge polarization and Majorana-Wannier states to the effective Floquet-Bloch eigenstates, we show that the TPTs in the driven case are signalled by a divergence of the correlation length of the Majorana-Wannier state correlation function.  A calculation of the  critical exponents reveals  that  TPTs in both the static and periodically driven  chain  belong to the same universality class.

The fixed lines of the CRG flow, on the other hand, reflect the frozen dynamics of the system, where the quasienergy dispersion maps back to the static dispersion. The fixed lines provide an
explanation for previously unexplained features seen   in    Majorana correlations in open systems.  Surprisingly,
some of the FLs indicate new topological instabilities where the total number of $0-$ and $\pi-$FMF's changes  by $2$.
Because of the simultaneous appearance of pairs of $0$ or $\pi$-FMF's, these transitions are speculated to be of the anomalous kind, across which the difference $|N_0 - N_{\pi}|$ stays constant.

An interesting question awaiting exploration is whether, for such intriguing systems, the CRG methodology captures the complexity of the topology stemming from the underlying micromotion.
In particular, given the preliminary results indicating additional transitions where $|\Delta{\cal M}| = 2$, it would be intriguing to apply the CRG method to 2D models that are known to host anomalous topological phases~\citep{Rudner:2013, Tauber:2018}.

We anticipate that our CRG method may be broadly applied to investigate TPTs and universality classes in Floquet systems subject to other types of periodic driving, such as square waves or multistep driving, or models defined in higher spatial dimensions.  Furthermore, the discretized RG equation, Eq.~(\ref{RG_eq_numerical}), offers a  very efficient numerical tool to identify TPTs especially in driven  higher dimensional systems.

%%%%%%%%%%%%%%%%%%
%%%     Acknowledgments    %%%
%%%%%%%%%%%%%%%%%%
\section{Acknowledgments}
We kindly acknowledge financial support by Giulio Anderheggen and the ETH Z\"{u}rich Foundation.
The authors would like to thank Manisha Thakurathi, Aline Ramires, Luca Papariello, Ivo Maceira, and Cl\'ement Tauber for fruitful discussions.

%%%%%%%%%%%%%%%
% %%    Appendices     %%%
%%%%%%%%%%%%%%%
\appendix

%%%%%%%%%%%%%%%%
% %%    Appendix A1       %%%
%%%%%%%%%%%%%%%%
\setcounter{section}{0}
\renewcommand{\thesection}{A\arabic{section}}
\section{Decay length of the Majorana edge state close to the static TPTs}
\label{appendix:decay_length_Majorana_static}

\renewcommand{\thesection}{A\arabic{section}.}

To show that the correlation length of the Wannier state correlation function coincides with the decay length of the Majorana edge state in the static Kitaev chain, we address the edge state explicitly. 
We first consider the edge state when the system is about to have a gap-closing at an HSP $k_{0}=\left\{0,\pi\right\}$. 
We aim at solving for the zero energy edge state satisfying $({\bf a}_{k}\cdot{\boldsymbol\tau})\psi(x)=0$ for a Hamiltonian defined in the positive half-space $x\geq 0$.
Expanding the Hamiltonian around $k_{0}$ to leading order and project the Hamiltonian into real space
by $k=-i\partial_{x}$ gives
\begin{eqnarray}
&&a_{2,k_{0}+k}=k\left(\partial_{k}a_{2,k}\right)_{k_{0}}=-i\left(\partial_{k}a_{2,k}\right)_{k_{0}}\partial_{x}\;,
\nonumber \\
&&a_{3,k_{0}+k}=a_{3,k_{0}}+\frac{k^{2}}{2}\left(\partial_{k}^{2}a_{3,k}\right)_{k_{0}}
\nonumber \\
&&=a_{3,k_{0}}-\frac{1}{2}\left(\partial_{k}^{2}a_{3,k}\right)_{k_{0}}\partial_{x}^{2}
\;,
\nonumber \\
&&\left\{-i\left(\partial_{k}a_{2,k}\right)_{k_{0}}\tau^{y}\partial_{x}\right.
\nonumber \\
&&\left.+\left[a_{3,k_{0}}-\frac{1}{2}\left(\partial_{k}^{2}a_{3,k}\right)_{k_{0}}\partial_{x}^{2}\right]\tau^{z}\right\}\psi(x)=0\;,
\nonumber \\
\label{edge_state_h_expansion}
\end{eqnarray}
where $\left(\partial_{k}a_{2,k}\right)_{k_{0}}$ denotes $\partial_{k}a_{2,k}$ evaluated at $k=k_{0}$. 
Multiplying the equation by $\tau^{y}$, we see that the edge state is an eigenstate of $\tau^{x}$, with ansatz 
$\psi=\chi_{\eta}\phi(x)\propto\chi_{\eta}e^{-x/\xi_{k_0}}$, where $\tau^{x}\chi_{\eta}=\eta\chi_{\eta}=\pm\chi_{\eta}$.

The solution for the decay length is 
\begin{eqnarray}
\xi_{k_0, \pm}^{-1}&=&\frac{(\partial_{k}a_{2,k})_{k_{0}}}{\eta (\partial_{k}^{2}a_{3,k})_{k_{0}}}
\nonumber \\
&&\pm \frac{|(\partial_{k}a_{2,k})_{k_{0}}|}{|\eta| (\partial_{k}^{2}a_{3,k})_{k_{0}}}
\sqrt{1+2\eta^2 \frac{a_{3,k}(\partial_{k}^{2}a_{3,k})_{k_{0}}}{(\partial_{k}a_{2,k})_{k_{0}}^{2}}}.\;\;\;
\end{eqnarray}
Demanding the sum of the two $\xi_{k_0,+}^{-1}+\xi_{k_0, -}^{-1}=2\eta(\partial_{k}a_{2,k})_{k_{0}}/(\partial_{k}^{2}a_{3,k})_{k_{0}}>0$
yields $\eta={\rm Sgn}\left[(\partial_{k}a_{2,k})_{k_{0}}/(\partial_{k}^{2}a_{3,k})_{k_{0}}\right]$.
There can be two cases: (a) If $(\partial_{k}^{2}a_{3,k})_{k_{0}}>0$, then $\eta={\rm Sgn}\left[(\partial_{k}a_{2,k})_{k_{0}}\right]$ and hence the longer one 
\begin{eqnarray}
\xi_{k_0,-}=-\frac{|(\partial_{k}a_{2,k})_{k_{0}}|}{a_{3,k_{0}}}
\end{eqnarray}
is identified as the decay length, where we have expanded the square root to obtain this solution.
The decay length must be positive, and $a_{3,k_{0}}<0$ must be satisfied in order for the edge state to exist. 
(b) If $(\partial_{k}^{2}a_{3,k})_{k_{0}}<0$, then $\eta=-{\rm Sgn}\left[(\partial_{k}a_{2,k})_{k_{0}}\right]$ and hence the longer one
\begin{eqnarray}
\xi_{k_0,+}=\frac{|(\partial_{k}a_{2,k})_{k_{0}}|}{a_{3,k_{0}}}
\end{eqnarray}
is the decay length.
Demanding it to be positive, one sees that only when $a_{3,k_{0}}>0$ does the edge state exist.
In summary, across the TPT $(\partial_{k}^{2}a_{3,k})_{k_{0}}a_{3,k_{0}}$ changes sign, and the edge state exists in the phase that has
\begin{eqnarray}
(\partial_{k}^{2}a_{3,k})_{k_{0}}a_{3,k_{0}}<0\;,
\end{eqnarray}
with a decay length 
\begin{eqnarray}
\xi_{k_0}=\left|\frac{(\partial_{k}a_{2,k})_{k_{0}}}{a_{3,k_{0}}}\right|\;.
\label{static_decay_length_general}
\end{eqnarray}
In the static Kitaev chain, using Eq.~(\ref{Dirac_model_static_Kitaev}) and $k_{0}=\left\{0,\pi\right\}$ yields
\begin{eqnarray}
&&k_{0}=0:\;{\rm exits\;when}\;\mu_{0}<t\;{\rm and}\;\xi_{0}=\left|\frac{\Delta}{t-\mu_0}\right|\;,
\nonumber \\
&&k_{0}=\pi:\;{\rm exits\;when}\;\mu_{0}>-t\;{\rm and}\;\xi_{\pi}=\left|\frac{\Delta}{t+\mu_0}\right|\;.
\nonumber \\
\end{eqnarray}
If $\eta=1$, then one may choose the spinor of the edge state to be $\chi_{\eta}=(1,1)^{T}/\sqrt{2}$, meaning that the edge state annihilation operator $\psi=\left(f+f^{\dag}\right)/\sqrt{2}=\psi^{\dag}$ is its own creation operator. 
Likewisely, if $\eta=-1$, then one may choose $\chi_{\eta}=(i,-i)^{T}/\sqrt{2}$ and hence the edge state annihilation operator $\psi=\left(if-if^{\dag}\right)/\sqrt{2}=\psi^{\dag}$ is again its own creation operator. 
Thus the annihilation operator of the edge state is a Majorana fermion.
Comparing the decay length $\xi_{k_0}$ with Eq.~(\ref{static_Kitaev_xi_Fk0}), it is evident that the decay length of the Majorana edge state that appears in the open boundary condition coincides with the correlation length $\xi$ of the Wannier state correlation function defined in the closed boundary condition.

%%%%%%%%%%%%%%%
%%%    Appendix A2     %%%
%%%%%%%%%%%%%%%
 \renewcommand{\thesection}{A\arabic{section}}
\section{Decay length of the Floquet-Majorana edge state close to the driven TPTs}
\label{appendix:decay_length_Majorana_driven}
\renewcommand{\thesection}{A\arabic{section}.}

The same analysis is also applicable to the periodically driven case, in which case we look for the localized zero Floquet energy edge state satisfying $h_{\text{eff},k}\psi=(\tilde{\bf a}_{k}\cdot{\boldsymbol\tau})\psi=0$, or equivalently from Eq.~(\ref{heff_definition}),
\begin{eqnarray}
\left[\tilde{a}_{2,k}\tau^{y}+\tilde{a}_{3,k}\tau^{z}\right]\psi(x)=0\;.
\label{driven_Dirac_real_space}
\end{eqnarray} 
Following the same calculation from Eq.~(\ref{edge_state_h_expansion}) to (\ref{static_decay_length_general}), we see that across the TPTs $(\partial_{k}^{2}\tilde{a}_{3,k})_{k_{0}}\tilde{a}_{3,k_{0}}$ changes sign, and the edge state exists in the phase that has
\begin{eqnarray}
(\partial_{k}^{2}\tilde{a}_{3,k})_{k_{0}}\tilde{a}_{3,k_{0}}<0\;,
\label{driven_Majorana_existence_criterion}
\end{eqnarray}
as we have verified numerically, and the decay length is
\begin{eqnarray}
\xi_{k_0}=\left|\frac{(\partial_{k}\tilde{a}_{2,k})_{k_{0}}}{\tilde{a}_{3,k_{0}}}\right|\;.
\label{driven_decay_length_general}
\end{eqnarray}
Comparing with Eq.~(\ref{driven_xi_Fk0_bk}), the correspondence between the decay length and the Majorana-Wannier state correlation length $\xi$ is evident.

We proceed to discuss the critical exponent of $\xi_{k_0}=\xi$ near the TPTs in the driven case. 
From Eq.~(\ref{heff_definition}), we see that at the HSP $k=\left\{0,\pi\right\}$, the $\tilde{a}_{2,k}$ vanishes at any $\left\{T,\mu_{1}\right\}$, so the gap-closing at $k_{0}=\left\{0,\pi\right\}$ is entirely determined by when the $\tilde{a}_{3,k}$ term vanishes.
First let us consider the TPT caused by tuning $T$ but holding $\mu_{1}$ fixed. The critical point $T_{c}$ thus satisfies $\tilde{a}_{3,k_{0},T_{c}}=0$.
Expand $\tilde{a}_{3,k}$ near the critical point $T_{c}$ yields
\begin{eqnarray}
\tilde{a}_{3,k_{0},T_{c}+\delta T} &&= \tilde{a}_{3,k_{0},T_{c}}+\delta T\;(\partial_{T}\tilde{a}_{3,k_{0},T})_{T_{c}}
\nonumber \\
&&=\delta T\;(\partial_{T}\tilde{a}_{3,k_{0},T})_{T_{c}}\;,
\end{eqnarray}
provided the leading order expansion does not vanish, which is true for this Floquet Majorana problem.
Therefore, the decay length in Eq.~(\ref{driven_decay_length_general}) near the critical point scales like
\begin{eqnarray}
\xi_{k_0}|_{T_{c}+\delta T}\propto\frac{1}{\delta T\;(\partial_{T}\tilde{a}_{3,k_{0},T})_{T_{c}}}\propto\frac{1}{T-T_{c}}\;,
\end{eqnarray}
indicating its critical exponent is $\nu=1$ when $T$ approaches $T_{c}$.
The same argument also holds when one varies $\mu_{1}$ across the critical point $\mu_{1c}$ holding $T$ fixed, in which case we expand
\begin{eqnarray}
\tilde{a}_{3,k_{0},\mu_{1c}+\delta\mu_{1}} &&=\tilde{a}_{3,k_{0},\mu_{1c}}+\delta \mu_{1}\;(\partial_{\mu_{1}}\tilde{a}_{3,k_{0},\mu_{1}})_{\mu_{1c}}
\nonumber \\
&&=\delta \mu_{1}\;(\partial_{\mu_{1}}\tilde{a}_{3,k_{0},\mu_{1}})_{\mu_{1c}}\;.
\end{eqnarray}
The decay length near the critical point scales like 
\begin{eqnarray}
\xi_{k_0}|_{\mu_{1c}+\delta\mu_{1}}\propto\frac{1}{\delta \mu_{1}\;(\partial_{\mu_{1}}\tilde{a}_{3,k_{0},\mu_{1}})_{\mu_{1c}}}\propto\frac{1}{\mu_{1}-\mu_{1c}}.
\end{eqnarray}
In short, whether approaching the phase boundary $\left\{T_{c},\mu_{1c}\right\}$ by varying $T$ or $\mu_{1}$, the critical exponent of the edge state decay length is $\nu=1$.

%%%%%%%%%%%%%%%
% %%    Appendix B      %%%
%%%%%%%%%%%%%%%
\setcounter{section}{1}
\renewcommand{\thesection}{\Alph{section}}
\section{Behavior of quasienergy spectrum and eigenfunctions across $|\Delta \mathcal{M}|=2$ transitions}
\label{appendix:figures_2Maj_jumps}

\begin{figure}[ht]
\centering
\includegraphics[width=\columnwidth]{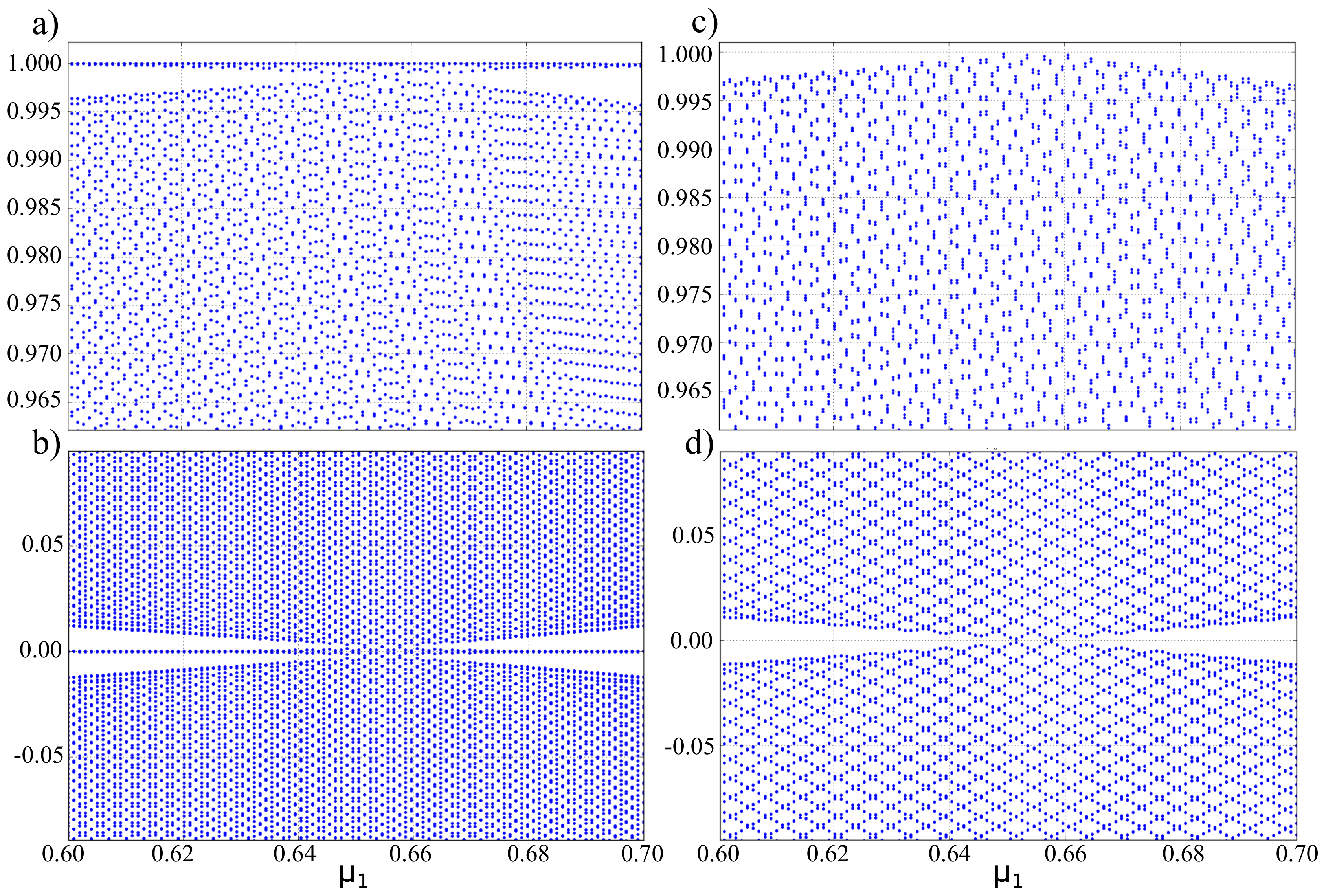}
\caption{Normalized quasienergy spectrum $\epsilon_i/\pi$ across the LCFL at $\mu_0=0.5$, $T=2.4$, $\mu_1=0.6-0.7$ for open boundary conditions (panels a) and b)) and periodic boundary conditions (panels c) and d)). The spectrum is shown in proximity of quasienergy 0 (panels b) and d)) and $\pi$ (panels a) and c)). A small but sizeable gap closes at $\mu_1 \approx 0.65$ in correspondence to the fixed point of the CRG flow, also for periodic boundary conditions.}
\label{figure:quasienergy-ribs}
\end{figure}

\begin{figure}[ht]
\centering
\includegraphics[width=\columnwidth]{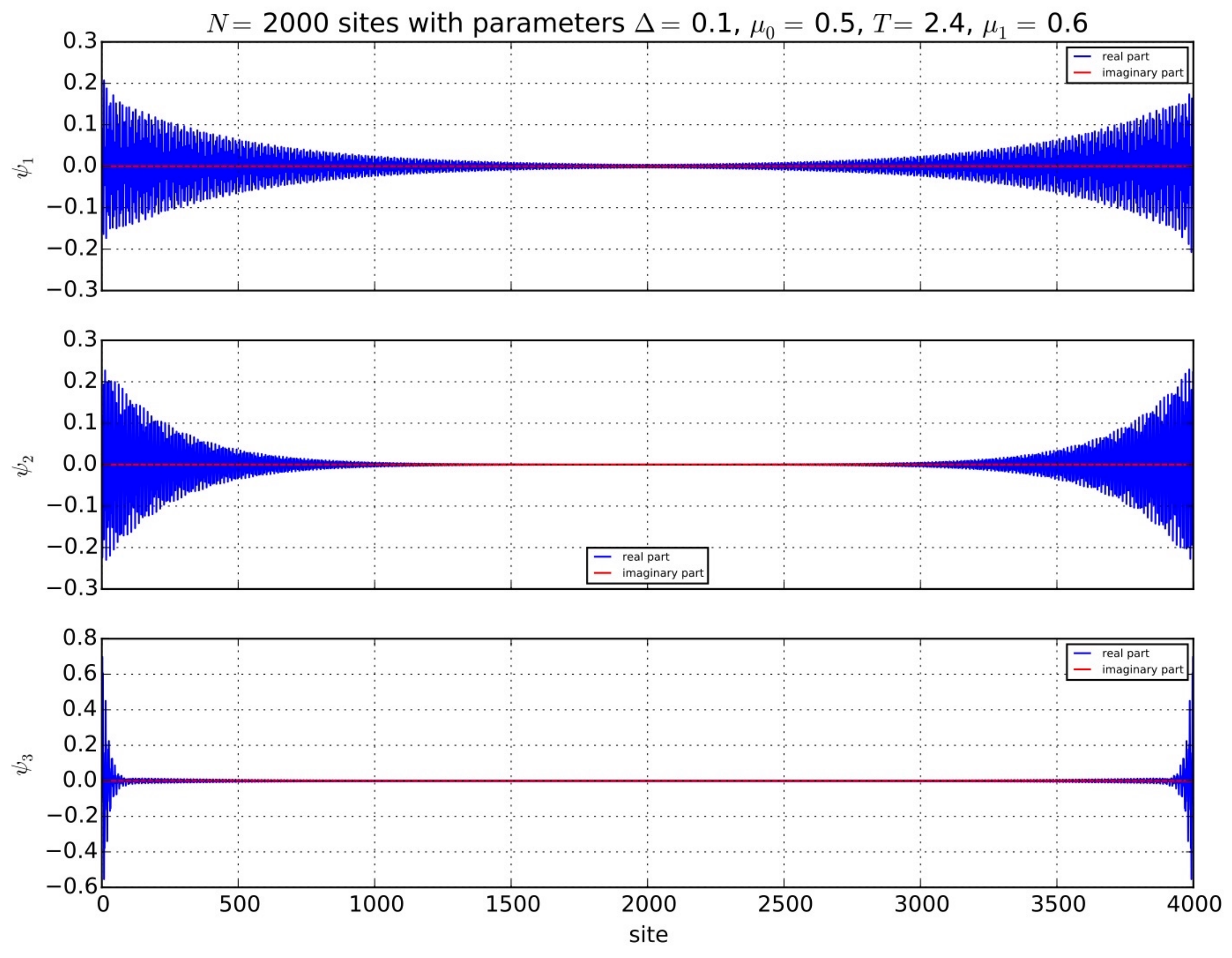}
\caption{Illustration of the eigenfunctions corresponding to the quasienergies $\pi$ ($\psi_1$ and $\psi_2$) and 0 ($\psi_3$) below the FL at $T=2.4$, $\mu_1=0.6$ for a chain of $N=2000$ fermions. The other parameters are $\Delta=0.1$ and $\mu_0=0.5$. Note that the additional $\pi$-modes tend to localize at the edges, but their localization length is much larger than the one of the $0$-mode (bottom panel). These modes are expected to localize asymptotically as $N \to \infty$.}
\label{figure:eigenfunctions-ribs}
\end{figure}

As explained in the main text, simultaneous divergences at non-HSP in the Berry connection can lead to the appearance of regions where the number of Floquet-Majorana modes jumps by two, \textit{i.e.} $|\Delta{\cal M}| = 2$.
To understand the character of these transitions we illustrate here some graphical results pertaining to these regions.
To verify whether the number of FMF's indeed changes across the FLs, we have calculated the quasienergy spectrum for chains up to $N=2000$ (see Fig. \ref{figure:quasienergy-ribs}). 
We find that in the 2-FMF region for $\mu_0=0.1$, two pairs of eigenvalues $\pm \epsilon_i$ approach $\pm \pi$ within $10^{-5}$, while a finite gap ($10^{-2}$) remains at 0. 
The eigenvalues close to $\pi$ are separated from the next eigenvalues by a finite gap of the same order. Similarly, in the 3-FMF region for $\mu_0=0.5$, a pair of eigenvalues approaches 0, while two pairs approach $\pm \pi$ within $10^{-5}$. 
They are again separated from the next eigenvalues by a gap of at least $10^{-2}$. 
We have confirmed that these eigenvalues converge to $0$ or $\pi$ as $N \to \infty$, while the gaps stay finite.
However, the relative magnitude of the gaps is too small to convincingly substantiate the appearance of isolated quasienergies at 0 and $\pm \pi$.
In order to understand the character of those additional asymptotic zero-energy Floquet modes, we have furthermore plotted their eigenvectors and discovered that they can always be chosen to be purely real.
Additionally, they appear to be localized at the edges, although their localization length, contrary to the FMF's obtained deep into the topological phases, can stretch over hundreds of sites and hence tend to hybridize them out of the 0- or $\pi$-energy (see Fig. \ref{figure:eigenfunctions-ribs}.)
The asymptotic eigenmodes seem to increase their localization in the limit $N \to \infty$. 
In summary, current results reach the limits of our numerical accuracy and we can therefore not conclusively confirm that the asymptotic zero-energy Floquet modes can be classified as true FMF's.
Typically,  the fixed lines do not manifest in the topological phase diagram derived from $\mathcal{M} = N_{0} + N_{\pi}$. 
However, for certain values of the static parameter $\mu_0$, parts of the lines are detected at higher periods $T \gtrsim 2.0$ as additional transitions where the value of $\mathcal{M}$ jumps by two (see \textit{e.g.} figure \ref{top-invariants2}). 
It is however unclear to us whether this principle remains in other Floquet systems or other types of driving potential.

%%%%%
%%%%%

%%%%%
%%%%%

%%%%%%%%%%%%%%%%%%%%%%
\bibliography{many-body-biblio}

\begin{thebibliography}{45}
\expandafter\ifx\csname natexlab\endcsname\relax\def\natexlab#1{#1}\fi
\expandafter\ifx\csname bibnamefont\endcsname\relax
  \def\bibnamefont#1{#1}\fi
\expandafter\ifx\csname bibfnamefont\endcsname\relax
  \def\bibfnamefont#1{#1}\fi
\expandafter\ifx\csname citenamefont\endcsname\relax
  \def\citenamefont#1{#1}\fi
\expandafter\ifx\csname url\endcsname\relax
  \def\url#1{\texttt{#1}}\fi
\expandafter\ifx\csname urlprefix\endcsname\relax\def\urlprefix{URL }\fi
\providecommand{\bibinfo}[2]{#2}
\providecommand{\eprint}[2][]{\url{#2}}

\bibitem[{\citenamefont{Wen}(1990)}]{Wen:1990}
\bibinfo{author}{\bibfnamefont{X.-G.} \bibnamefont{Wen}},
  \bibinfo{journal}{Int. J. Mod. Phys.} p. \bibinfo{pages}{239}
  (\bibinfo{year}{1990}).

\bibitem[{\citenamefont{Landau}(1937)}]{Landau}
\bibinfo{author}{\bibfnamefont{L.~D.} \bibnamefont{Landau}},
  \bibinfo{journal}{Zh. Eksp. Teor. Fiz.} \textbf{\bibinfo{volume}{7}},
  \bibinfo{pages}{19} (\bibinfo{year}{1937}).

\bibitem[{\citenamefont{Miransky}(1994)}]{Miransky-book}
\bibinfo{author}{\bibfnamefont{V.~A.} \bibnamefont{Miransky}},
  \emph{\bibinfo{title}{Dynamical Symmetry Breaking in Quantum Field Theories}}
  (\bibinfo{publisher}{World Scientific Publishing Co.}, \bibinfo{year}{1994}).

\bibitem[{\citenamefont{Thouless et~al.}(1982)\citenamefont{Thouless, Kohmoto,
  Nightingale, and den Nijs}}]{Thouless:1982}
\bibinfo{author}{\bibfnamefont{D.~J.} \bibnamefont{Thouless}},
  \bibinfo{author}{\bibfnamefont{M.}~\bibnamefont{Kohmoto}},
  \bibinfo{author}{\bibfnamefont{M.~P.} \bibnamefont{Nightingale}},
  \bibnamefont{and} \bibinfo{author}{\bibfnamefont{M.}~\bibnamefont{den Nijs}},
  \bibinfo{journal}{Phys. Rev. Lett.} p. \bibinfo{pages}{405}
  (\bibinfo{year}{1982}).

\bibitem[{\citenamefont{Wen}(1989)}]{Wen:1989}
\bibinfo{author}{\bibfnamefont{X.-G.} \bibnamefont{Wen}},
  \bibinfo{journal}{Phys. Rev. B} p. \bibinfo{pages}{7387}
  (\bibinfo{year}{1989}).

\bibitem[{\citenamefont{Qi et~al.}(2008)\citenamefont{Qi, Hughes, and
  Zhang}}]{Qi:2008}
\bibinfo{author}{\bibfnamefont{X.-L.} \bibnamefont{Qi}},
  \bibinfo{author}{\bibfnamefont{T.}~\bibnamefont{Hughes}}, \bibnamefont{and}
  \bibinfo{author}{\bibfnamefont{S.-C.} \bibnamefont{Zhang}},
  \bibinfo{journal}{Phys. Rev. B} p. \bibinfo{pages}{195424}
  (\bibinfo{year}{2008}).

\bibitem[{\citenamefont{Wen}(1995)}]{Wen:1995}
\bibinfo{author}{\bibfnamefont{X.-G.} \bibnamefont{Wen}},
  \bibinfo{journal}{Advances in Physics} p. \bibinfo{pages}{405}
  (\bibinfo{year}{1995}).

\bibitem[{\citenamefont{Chen et~al.}(2010)\citenamefont{Chen, Gu, and
  Wen}}]{XieChen:2010}
\bibinfo{author}{\bibfnamefont{X.}~\bibnamefont{Chen}},
  \bibinfo{author}{\bibfnamefont{Z.-C.} \bibnamefont{Gu}}, \bibnamefont{and}
  \bibinfo{author}{\bibfnamefont{X.-G.} \bibnamefont{Wen}},
  \bibinfo{journal}{Phys. Rev. B} p. \bibinfo{pages}{155138}
  (\bibinfo{year}{2010}).

\bibitem[{\citenamefont{Fidkowski}(2010)}]{Fidkowski:2010}
\bibinfo{author}{\bibfnamefont{L.}~\bibnamefont{Fidkowski}},
  \bibinfo{journal}{Phys. Rev. Lett.} p. \bibinfo{pages}{130502}
  (\bibinfo{year}{2010}).

\bibitem[{\citenamefont{Wen}(1991)}]{Wen:1991}
\bibinfo{author}{\bibfnamefont{X.-G.} \bibnamefont{Wen}},
  \bibinfo{journal}{Phys. Rev. B} p. \bibinfo{pages}{11025}
  (\bibinfo{year}{1991}).

\bibitem[{\citenamefont{Kane and Mele}(2005)}]{Kane-Mele2005}
\bibinfo{author}{\bibfnamefont{C.~L.} \bibnamefont{Kane}} \bibnamefont{and}
  \bibinfo{author}{\bibfnamefont{E.~J.} \bibnamefont{Mele}},
  \textbf{\bibinfo{volume}{95 (14)}}, \bibinfo{pages}{146802}
  (\bibinfo{year}{2005}).

\bibitem[{\citenamefont{Goldman and Su}(1995)}]{Goldman:1995}
\bibinfo{author}{\bibfnamefont{V.~J.} \bibnamefont{Goldman}} \bibnamefont{and}
  \bibinfo{author}{\bibfnamefont{B.}~\bibnamefont{Su}},
  \bibinfo{journal}{Science} p. \bibinfo{pages}{1010} (\bibinfo{year}{1995}).

\bibitem[{\citenamefont{de~Picciotto et~al.}(1997)\citenamefont{de~Picciotto,
  Reznikov, Heiblum, Umansky, Bunin, and Mahalu}}]{dePicciotto:1997}
\bibinfo{author}{\bibfnamefont{R.}~\bibnamefont{de~Picciotto}},
  \bibinfo{author}{\bibfnamefont{M.}~\bibnamefont{Reznikov}},
  \bibinfo{author}{\bibfnamefont{M.}~\bibnamefont{Heiblum}},
  \bibinfo{author}{\bibfnamefont{V.}~\bibnamefont{Umansky}},
  \bibinfo{author}{\bibfnamefont{G.}~\bibnamefont{Bunin}}, \bibnamefont{and}
  \bibinfo{author}{\bibfnamefont{D.}~\bibnamefont{Mahalu}},
  \bibinfo{journal}{Nature} p. \bibinfo{pages}{162} (\bibinfo{year}{1997}).

\bibitem[{\citenamefont{Martin et~al.}(2004)\citenamefont{Martin, Ilani,
  Verdene, Smet, Umansky, Mahalu, Schuh, Abstreiter, and Yacoby}}]{Martin:2004}
\bibinfo{author}{\bibfnamefont{J.}~\bibnamefont{Martin}},
  \bibinfo{author}{\bibfnamefont{S.}~\bibnamefont{Ilani}},
  \bibinfo{author}{\bibfnamefont{B.}~\bibnamefont{Verdene}},
  \bibinfo{author}{\bibfnamefont{J.}~\bibnamefont{Smet}},
  \bibinfo{author}{\bibfnamefont{V.}~\bibnamefont{Umansky}},
  \bibinfo{author}{\bibfnamefont{D.}~\bibnamefont{Mahalu}},
  \bibinfo{author}{\bibfnamefont{D.}~\bibnamefont{Schuh}},
  \bibinfo{author}{\bibfnamefont{G.}~\bibnamefont{Abstreiter}},
  \bibnamefont{and} \bibinfo{author}{\bibfnamefont{A.}~\bibnamefont{Yacoby}},
  \bibinfo{journal}{Science} pp. \bibinfo{pages}{980--3}
  (\bibinfo{year}{2004}).

\bibitem[{\citenamefont{Moore and Read}(1991)}]{Moore}
\bibinfo{author}{\bibfnamefont{G.}~\bibnamefont{Moore}} \bibnamefont{and}
  \bibinfo{author}{\bibfnamefont{N.}~\bibnamefont{Read}},
  \bibinfo{journal}{Nucl. Phys.} \textbf{\bibinfo{volume}{B360}},
  \bibinfo{pages}{362} (\bibinfo{year}{1991}).

\bibitem[{\citenamefont{Kitaev}(2001)}]{Kitaev2001}
\bibinfo{author}{\bibfnamefont{A.~Y.} \bibnamefont{Kitaev}},
  \bibinfo{journal}{Usp. Fiz. Nauk.} \textbf{\bibinfo{volume}{171 (10)}},
  \bibinfo{pages}{131} (\bibinfo{year}{2001}).

\bibitem[{\citenamefont{Mong et~al.}(2014)\citenamefont{Mong, Clarke, Alicea,
  Lindner, Fendley, Nayak, Oreg, Stern, Berg, Shtengel et~al.}}]{Mong:2014}
\bibinfo{author}{\bibfnamefont{R.~S.~K.} \bibnamefont{Mong}},
  \bibinfo{author}{\bibfnamefont{D.~J.} \bibnamefont{Clarke}},
  \bibinfo{author}{\bibfnamefont{J.}~\bibnamefont{Alicea}},
  \bibinfo{author}{\bibfnamefont{N.~H.} \bibnamefont{Lindner}},
  \bibinfo{author}{\bibfnamefont{P.}~\bibnamefont{Fendley}},
  \bibinfo{author}{\bibfnamefont{C.}~\bibnamefont{Nayak}},
  \bibinfo{author}{\bibfnamefont{Y.}~\bibnamefont{Oreg}},
  \bibinfo{author}{\bibfnamefont{A.}~\bibnamefont{Stern}},
  \bibinfo{author}{\bibfnamefont{E.}~\bibnamefont{Berg}},
  \bibinfo{author}{\bibfnamefont{K.}~\bibnamefont{Shtengel}},
  \bibnamefont{et~al.}, \bibinfo{journal}{Phys. Rev. X} p.
  \bibinfo{pages}{011036} (\bibinfo{year}{2014}).

\bibitem[{\citenamefont{Sato and Saitoh}(2015)}]{Sato-Spintronics-book}
\bibinfo{author}{\bibfnamefont{K.}~\bibnamefont{Sato}} \bibnamefont{and}
  \bibinfo{author}{\bibfnamefont{E.}~\bibnamefont{Saitoh}},
  \emph{\bibinfo{title}{Spintronics for Next Generation Innovative Devices}},
  Wiley Materials for Electronic and Optoelectronic Applications
  (\bibinfo{publisher}{Wiley}, \bibinfo{year}{2015}).

\bibitem[{\citenamefont{Dennis et~al.}(2002)\citenamefont{Dennis, Kitaev,
  Landahl, and Preskill}}]{Dennis:2002}
\bibinfo{author}{\bibfnamefont{E.}~\bibnamefont{Dennis}},
  \bibinfo{author}{\bibfnamefont{A.}~\bibnamefont{Kitaev}},
  \bibinfo{author}{\bibfnamefont{A.}~\bibnamefont{Landahl}}, \bibnamefont{and}
  \bibinfo{author}{\bibfnamefont{J.}~\bibnamefont{Preskill}},
  \bibinfo{journal}{J. Math. Phys.} p. \bibinfo{pages}{4452}
  (\bibinfo{year}{2002}).

\bibitem[{\citenamefont{Kitaev}(2003)}]{Kitaev:2003}
\bibinfo{author}{\bibfnamefont{A.~Y.} \bibnamefont{Kitaev}},
  \bibinfo{journal}{Ann. Phys.} \textbf{\bibinfo{volume}{302}},
  \bibinfo{pages}{2} (\bibinfo{year}{2003}).

\bibitem[{\citenamefont{Nayak et~al.}(2008)\citenamefont{Nayak, Simon, Stern,
  Freedman, and Sarma}}]{Nayak:2008}
\bibinfo{author}{\bibfnamefont{C.}~\bibnamefont{Nayak}},
  \bibinfo{author}{\bibfnamefont{S.~H.} \bibnamefont{Simon}},
  \bibinfo{author}{\bibfnamefont{A.}~\bibnamefont{Stern}},
  \bibinfo{author}{\bibfnamefont{M.}~\bibnamefont{Freedman}}, \bibnamefont{and}
  \bibinfo{author}{\bibfnamefont{S.~D.} \bibnamefont{Sarma}},
  \bibinfo{journal}{Rev. Mod. Phys.} p. \bibinfo{pages}{1083}
  (\bibinfo{year}{2008}).

\bibitem[{\citenamefont{Lindner et~al.}(2011)\citenamefont{Lindner, Refael, and
  Galitski}}]{Lindner}
\bibinfo{author}{\bibfnamefont{N.~H.} \bibnamefont{Lindner}},
  \bibinfo{author}{\bibfnamefont{G.}~\bibnamefont{Refael}}, \bibnamefont{and}
  \bibinfo{author}{\bibfnamefont{V.}~\bibnamefont{Galitski}},
  \bibinfo{journal}{Nat. Phys.} \textbf{\bibinfo{volume}{7}},
  \bibinfo{pages}{490} (\bibinfo{year}{2011}).

\bibitem[{\citenamefont{Kitagawa et~al.}(2010)\citenamefont{Kitagawa, Berg,
  Rudner, and Demler}}]{Kitagawa}
\bibinfo{author}{\bibfnamefont{T.}~\bibnamefont{Kitagawa}},
  \bibinfo{author}{\bibfnamefont{E.}~\bibnamefont{Berg}},
  \bibinfo{author}{\bibfnamefont{M.}~\bibnamefont{Rudner}}, \bibnamefont{and}
  \bibinfo{author}{\bibfnamefont{E.}~\bibnamefont{Demler}},
  \bibinfo{journal}{Phys. Rev. B} \textbf{\bibinfo{volume}{82}},
  \bibinfo{pages}{235114} (\bibinfo{year}{2010}).

\bibitem[{\citenamefont{Liu et~al.}(2013)\citenamefont{Liu, Levchenko, and
  Baranger}}]{Liu}
\bibinfo{author}{\bibfnamefont{D.~E.} \bibnamefont{Liu}},
  \bibinfo{author}{\bibfnamefont{A.}~\bibnamefont{Levchenko}},
  \bibnamefont{and} \bibinfo{author}{\bibfnamefont{H.~U.}
  \bibnamefont{Baranger}}, \bibinfo{journal}{Phys. Rev. Lett.}
  \textbf{\bibinfo{volume}{111}}, \bibinfo{pages}{047002}
  (\bibinfo{year}{2013}).

\bibitem[{\citenamefont{Cayssol et~al.}(2013)\citenamefont{Cayssol, D{\'o}ra,
  Simon, and Moessner}}]{Cayssol:2013}
\bibinfo{author}{\bibfnamefont{J.}~\bibnamefont{Cayssol}},
  \bibinfo{author}{\bibfnamefont{B.}~\bibnamefont{D{\'o}ra}},
  \bibinfo{author}{\bibfnamefont{F.}~\bibnamefont{Simon}}, \bibnamefont{and}
  \bibinfo{author}{\bibfnamefont{R.}~\bibnamefont{Moessner}},
  \bibinfo{journal}{Phys. Status Solidi RRL} \textbf{\bibinfo{volume}{7}},
  \bibinfo{pages}{101} (\bibinfo{year}{2013}).

\bibitem[{\citenamefont{Harper and Roy}(2017)}]{Harper:2017}
\bibinfo{author}{\bibfnamefont{F.}~\bibnamefont{Harper}} \bibnamefont{and}
  \bibinfo{author}{\bibfnamefont{R.}~\bibnamefont{Roy}},
  \bibinfo{journal}{Phys. Rev. Lett.} \textbf{\bibinfo{volume}{118}},
  \bibinfo{pages}{115301} (\bibinfo{year}{2017}).

\bibitem[{\citenamefont{Roy and Harper}(2017)}]{Roy:2017}
\bibinfo{author}{\bibfnamefont{R.}~\bibnamefont{Roy}} \bibnamefont{and}
  \bibinfo{author}{\bibfnamefont{F.}~\bibnamefont{Harper}},
  \bibinfo{journal}{Phys. Rev. B} \textbf{\bibinfo{volume}{96}},
  \bibinfo{pages}{155118} (\bibinfo{year}{2017}).

\bibitem[{\citenamefont{Yao et~al.}(2017)\citenamefont{Yao, Yan, and
  Wang}}]{Yao:2017}
\bibinfo{author}{\bibfnamefont{S.}~\bibnamefont{Yao}},
  \bibinfo{author}{\bibfnamefont{Z.}~\bibnamefont{Yan}}, \bibnamefont{and}
  \bibinfo{author}{\bibfnamefont{Z.}~\bibnamefont{Wang}},
  \bibinfo{journal}{Phys. Rev. B} \textbf{\bibinfo{volume}{96}},
  \bibinfo{pages}{195303} (\bibinfo{year}{2017}).

\bibitem[{\citenamefont{Thakurathi et~al.}(2013)\citenamefont{Thakurathi,
  Patel, Sen, and Dutta}}]{Thakurathi}
\bibinfo{author}{\bibfnamefont{M.}~\bibnamefont{Thakurathi}},
  \bibinfo{author}{\bibfnamefont{A.~A.} \bibnamefont{Patel}},
  \bibinfo{author}{\bibfnamefont{D.}~\bibnamefont{Sen}}, \bibnamefont{and}
  \bibinfo{author}{\bibfnamefont{A.}~\bibnamefont{Dutta}},
  \bibinfo{journal}{Phys. Rev. B} \textbf{\bibinfo{volume}{88}},
  \bibinfo{pages}{155133} (\bibinfo{year}{2013}).

\bibitem[{\citenamefont{Molignini et~al.}(2017)\citenamefont{Molignini, van
  Nieuwenburg, and Chitra}}]{Molignini:2017}
\bibinfo{author}{\bibfnamefont{P.}~\bibnamefont{Molignini}},
  \bibinfo{author}{\bibfnamefont{E.}~\bibnamefont{van Nieuwenburg}},
  \bibnamefont{and} \bibinfo{author}{\bibfnamefont{R.}~\bibnamefont{Chitra}},
  \bibinfo{journal}{Phys. Rev. B}  (\bibinfo{year}{2017}).

\bibitem[{\citenamefont{Graf and Porta}(2013)}]{Graf:2013}
\bibinfo{author}{\bibfnamefont{G.~M.} \bibnamefont{Graf}} \bibnamefont{and}
  \bibinfo{author}{\bibfnamefont{M.}~\bibnamefont{Porta}},
  \bibinfo{journal}{Commun. Math. Phys.} p.~\bibinfo{pages}{85}
  (\bibinfo{year}{2013}).

\bibitem[{\citenamefont{Chiu et~al.}(2016)\citenamefont{Chiu, Teo, Schnyder,
  and Ryu}}]{ChiuReview:2016}
\bibinfo{author}{\bibfnamefont{C.-K.} \bibnamefont{Chiu}},
  \bibinfo{author}{\bibfnamefont{J.~C.~Y.} \bibnamefont{Teo}},
  \bibinfo{author}{\bibfnamefont{A.~P.} \bibnamefont{Schnyder}},
  \bibnamefont{and} \bibinfo{author}{\bibfnamefont{S.}~\bibnamefont{Ryu}},
  \bibinfo{journal}{Rev. Mod. Phys.} \textbf{\bibinfo{volume}{88}},
  \bibinfo{pages}{035055} (\bibinfo{year}{2016}).

\bibitem[{\citenamefont{Chen}(2016)}]{Chen:2016}
\bibinfo{author}{\bibfnamefont{W.}~\bibnamefont{Chen}}, \bibinfo{journal}{J.
  Phys.: Condens. Matter} \textbf{\bibinfo{volume}{28}},
  \bibinfo{pages}{055601} (\bibinfo{year}{2016}).

\bibitem[{\citenamefont{Chen et~al.}(2016)\citenamefont{Chen, Sigrist, and
  Schnyder}}]{Chen-Sigrist:2016}
\bibinfo{author}{\bibfnamefont{W.}~\bibnamefont{Chen}},
  \bibinfo{author}{\bibfnamefont{M.}~\bibnamefont{Sigrist}}, \bibnamefont{and}
  \bibinfo{author}{\bibfnamefont{A.~P.} \bibnamefont{Schnyder}},
  \bibinfo{journal}{J. Phys.: Condens. Matter} \textbf{\bibinfo{volume}{28}},
  \bibinfo{pages}{365501} (\bibinfo{year}{2016}).

\bibitem[{\citenamefont{Kourtis et~al.}(2017)\citenamefont{Kourtis, Neupert,
  Mudry, Sigrist, and Chen}}]{Kourtis:2017}
\bibinfo{author}{\bibfnamefont{S.}~\bibnamefont{Kourtis}},
  \bibinfo{author}{\bibfnamefont{T.}~\bibnamefont{Neupert}},
  \bibinfo{author}{\bibfnamefont{C.}~\bibnamefont{Mudry}},
  \bibinfo{author}{\bibfnamefont{M.}~\bibnamefont{Sigrist}}, \bibnamefont{and}
  \bibinfo{author}{\bibfnamefont{W.}~\bibnamefont{Chen}},
  \bibinfo{journal}{Phys. Rev. B} p. \bibinfo{pages}{205117}
  (\bibinfo{year}{2017}).

\bibitem[{\citenamefont{Chen}(2018)}]{Chen:2018}
\bibinfo{author}{\bibfnamefont{W.}~\bibnamefont{Chen}}, \bibinfo{journal}{Phys.
  Rev. B} \textbf{\bibinfo{volume}{97}}, \bibinfo{pages}{115130}
  (\bibinfo{year}{2018}),
  \urlprefix\url{https://link.aps.org/doi/10.1103/PhysRevB.97.115130}.

\bibitem[{\citenamefont{Rudner et~al.}(2013)\citenamefont{Rudner, Lindner,
  Berg, and Levin}}]{Rudner:2013}
\bibinfo{author}{\bibfnamefont{M.~S.} \bibnamefont{Rudner}},
  \bibinfo{author}{\bibfnamefont{N.~H.} \bibnamefont{Lindner}},
  \bibinfo{author}{\bibfnamefont{E.}~\bibnamefont{Berg}}, \bibnamefont{and}
  \bibinfo{author}{\bibfnamefont{M.}~\bibnamefont{Levin}},
  \bibinfo{journal}{Phys. Rev. X} \textbf{\bibinfo{volume}{3}},
  \bibinfo{pages}{031005} (\bibinfo{year}{2013}).

\bibitem[{\citenamefont{Chen et~al.}(2017)\citenamefont{Chen, Legner,
  R\"{u}egg, and Sigrist}}]{Chen:2017}
\bibinfo{author}{\bibfnamefont{W.}~\bibnamefont{Chen}},
  \bibinfo{author}{\bibfnamefont{M.}~\bibnamefont{Legner}},
  \bibinfo{author}{\bibfnamefont{A.}~\bibnamefont{R\"{u}egg}},
  \bibnamefont{and} \bibinfo{author}{\bibfnamefont{M.}~\bibnamefont{Sigrist}},
  \bibinfo{journal}{Phys. Rev. B} \textbf{\bibinfo{volume}{95}},
  \bibinfo{pages}{075116} (\bibinfo{year}{2017}).

\bibitem[{\citenamefont{King-Smith and Vanderbilt}(1993)}]{King-Smith:1993}
\bibinfo{author}{\bibfnamefont{R.~D.} \bibnamefont{King-Smith}}
  \bibnamefont{and}
  \bibinfo{author}{\bibfnamefont{D.}~\bibnamefont{Vanderbilt}},
  \bibinfo{journal}{Phys. Rev. B} \textbf{\bibinfo{volume}{47}},
  \bibinfo{pages}{1651} (\bibinfo{year}{1993}),
  \urlprefix\url{https://link.aps.org/doi/10.1103/PhysRevB.47.1651}.

\bibitem[{\citenamefont{Resta}(1994)}]{Resta:1994}
\bibinfo{author}{\bibfnamefont{R.}~\bibnamefont{Resta}}, \bibinfo{journal}{Rev.
  Mod. Phys.} \textbf{\bibinfo{volume}{66}}, \bibinfo{pages}{899}
  (\bibinfo{year}{1994}),
  \urlprefix\url{https://link.aps.org/doi/10.1103/RevModPhys.66.899}.

\bibitem[{\citenamefont{Dittrich et~al.}(1998)\citenamefont{Dittrich,
  H{\"a}nggi, Ingold, Kramer, Sch{\"o}n, and Zwerger}}]{Haenggi}
\bibinfo{author}{\bibfnamefont{T.}~\bibnamefont{Dittrich}},
  \bibinfo{author}{\bibfnamefont{P.}~\bibnamefont{H{\"a}nggi}},
  \bibinfo{author}{\bibfnamefont{G.-L.} \bibnamefont{Ingold}},
  \bibinfo{author}{\bibfnamefont{B.}~\bibnamefont{Kramer}},
  \bibinfo{author}{\bibfnamefont{G.}~\bibnamefont{Sch{\"o}n}},
  \bibnamefont{and} \bibinfo{author}{\bibfnamefont{W.}~\bibnamefont{Zwerger}},
  \emph{\bibinfo{title}{Quantum Transport and Dissipation}}
  (\bibinfo{publisher}{Wiley-VCH}, \bibinfo{year}{1998}).

\bibitem[{\citenamefont{Kitaev}(2009.)}]{Kitaev-table}
\bibinfo{author}{\bibfnamefont{A.}~\bibnamefont{Kitaev}}, in
  \emph{\bibinfo{booktitle}{AIP Conference Proceedings}}
  (\bibinfo{year}{2009.}), vol. \bibinfo{volume}{1134(1)}, pp.
  \bibinfo{pages}{22--30}.

\bibitem[{\citenamefont{Ezawa et~al.}(2013)\citenamefont{Ezawa, Tanaka, and
  Nagaosa}}]{Ezawa:2013}
\bibinfo{author}{\bibfnamefont{M.}~\bibnamefont{Ezawa}},
  \bibinfo{author}{\bibfnamefont{Y.}~\bibnamefont{Tanaka}}, \bibnamefont{and}
  \bibinfo{author}{\bibfnamefont{N.}~\bibnamefont{Nagaosa}},
  \bibinfo{journal}{Scientific Reports} \textbf{\bibinfo{volume}{3}}
  (\bibinfo{year}{2013}).

\bibitem[{\citenamefont{Prosen and Ilievski}(2011)}]{Prosen2011}
\bibinfo{author}{\bibfnamefont{T.}~\bibnamefont{Prosen}} \bibnamefont{and}
  \bibinfo{author}{\bibfnamefont{E.}~\bibnamefont{Ilievski}},
  \bibinfo{journal}{Phys. Rev. Lett.} \textbf{\bibinfo{volume}{107}},
  \bibinfo{pages}{060403} (\bibinfo{year}{2011}).

\bibitem[{\citenamefont{Graf and Tauber}(2018)}]{Tauber:2018}
\bibinfo{author}{\bibfnamefont{G.~M.} \bibnamefont{Graf}} \bibnamefont{and}
  \bibinfo{author}{\bibfnamefont{C.}~\bibnamefont{Tauber}},
  \bibinfo{journal}{Ann. Henri Poincar\ ́e} pp. \bibinfo{pages}{709--741}
  (\bibinfo{year}{2018}).

\end{thebibliography}
%%%%%%%%%%%%%%%%%%%%%%

\end{document}